\documentclass[11pt]{amsart}
\reversemarginpar
\newcommand{\commentout}[1]{}

\usepackage{amsmath}
\usepackage{amssymb}

\newcommand{\nwc}{\newcommand}

\newcommand{\cltil}{\tilde{\cL}^\ep}

\newcommand{\lt}{\left}
\newcommand{\rt}{\right}
\nwc{\ytil}{\tilde{\by}}
\nwc{\al}{\alpha}
\newcommand{\vas}{\varepsilon}
\newcommand{\lan}{\left\langle}
\newcommand{\ran}{\right\rangle}
\newcommand{\tvas}{W_z^\vas}
\newcommand{\psiep}{W_z^\vas}
\newcommand{\wep}{W^\vas}
\newcommand{\xtil}{\tilde{\bx}}
\newcommand{\xvec}{\vec{\bx}}
\newcommand{\kvec}{\vec{\bk}}

\newcommand{\wepz}{W_z^\vas}

\newcommand{\eptil}{\tilde{\ep}}
\newcommand{\vep}{{V}_z^\vas}
\newcommand{\cv}{{\ml L}^\ep_z}
\newcommand{\cvtil}{\tilde{{\ml L}}^\ep_z}
\newcommand{\vtil}{\tilde{V}^\ep}

\newcommand{\ks}{{k}}
\newcommand{\bx}{\mathbf x}
\newcommand{\bD}{\mathbf D}
\newcommand{\bp}{\mathbf p}
\newcommand{\br}{\mathbf r}
\newcommand{\bq}{\mathbf q}
\newcommand{\by}{\mathbf y}

\newcommand{\pdgx}{\bp\cdot\nabla_\bx}

\nwc{\nwt}{\newtheorem}

\nwt{prop}{Proposition}
\nwt{proposition}{Proposition}
\nwt{lemma}{Lemma}
\nwt{theorem}{Theorem}
\nwt{cor}{Corollary}
\nwt{corollary}{Corollary}
\nwt{remark}{Remark}
\nwt{assumption}{Assumption}
\nwt{definition}{Definition} 

\nwc{\bal}{\begin{align}}
\nwc{\be}{\begin{equation}}
\nwc{\ben}{\begin{equation*}}
\nwc{\bea}{\begin{eqnarray}}
\nwc{\beq}{\begin{eqnarray}}
\nwc{\bean}{\begin{eqnarray*}}
\nwc{\beqn}{\begin{eqnarray*}}
\nwc{\beqast}{\begin{eqnarray*}}

\nwc{\eal}{\end{align}}
\nwc{\ee}{\end{equation}}
\nwc{\een}{\end{equation*}}
\nwc{\eea}{\end{eqnarray}}
\nwc{\eeq}{\end{eqnarray}}
\nwc{\eean}{\end{eqnarray*}}
\nwc{\eeqn}{\end{eqnarray*}}
\nwc{\eeqast}{\end{eqnarray*}}

\nwc{\invf}{\cF^{-1}_2}
\nwc{\ep}{\varepsilon}
\nwc{\tep}{\tilde{\varepsilon}}
\nwc{\epsq}{{\varepsilon^2}}
\nwc{\epsqa}{{\varepsilon^{2\alpha}}}
\nwc{\eps}{\varepsilon}
\nwc{\ept}{\epsilon}
\nwc{\vrho}{\varrho}
\nwc{\orho}{\bar\varrho}
\nwc{\ou}{\bar u}
\nwc{\vpsi}{\varpsi}
\nwc{\lamb}{\lambda}

\nwc{\nn}{\nonumber}
\nwc{\bm}{\boldmath}
\nwc{\mf}{\mathbf}
\nwc{\mb}{\mathbf}
\nwc{\ml}{\mathcal}

\nwc{\IA}{\mathbb{A}} 
\nwc{\IB}{\mathbb{B}}
\nwc{\IC}{\mathbb{C}} 
\nwc{\ID}{\mathbb{D}} 
\nwc{\IM}{\mathbb{M}} 
\nwc{\IP}{\mathbb{P}} 
\nwc{\II}{\mathbb{I}} 
\nwc{\IE}{\mathbb{E}} 
\nwc{\IF}{\mathbb{F}} 
\nwc{\IG}{\mathbb{G}} 
\nwc{\IN}{\mathbb{N}} 
\nwc{\IQ}{\mathbb{Q}} 
\nwc{\IR}{\mathbb{R}} 
\nwc{\IT}{\mathbb{T}} 
\nwc{\IZ}{\mathbb{Z}} 

\nwc{\epal}{\ep^{-2\alpha}}
\nwc{\cE}{{\ml E}}
\nwc{\cP}{{\ml P}}
\nwc{\cQ}{{\ml Q}}
\nwc{\cL}{{\ml L}}
\nwc{\cR}{{\ml R}}
\nwc{\cV}{{\ml L}}
\nwc{\cT}{{\ml T}}
\nwc{\crV}{{\ml L}_{(\delta,\rho)}}
\nwc{\cC}{{\ml C}}
\nwc{\cA}{{\ml A}}
\nwc{\cK}{{\ml K}}
\nwc{\cB}{{\ml B}}
\nwc{\cD}{{\ml D}}
\nwc{\cF}{{\ml F}}
\nwc{\cS}{{\ml S}}
\nwc{\cM}{{\ml M}}
\nwc{\cG}{{\ml G}}
\nwc{\cH}{{\ml H}}
\nwc{\bk}{{\mb k}}
\nwc{\cbz}{\overline{\cB}_z}

\nwc{\pft}{\cF^{-1}_\bp}
\newcommand{\grx}{\nabla_\bx}
\newcommand{\gry}{\nabla_\by}
\newcommand{\grp}{\nabla_\bp}
\newcommand{\grxtil}{\nabla_{\tilde\bx}}
\begin{document}

%
\title{Self-Averaged Scaling Limits for
Random Parabolic Waves}

\author{Albert C. Fannjiang
 }
\thanks{Department of Mathematics,
University of California,
Davis, CA 95616
Internet: cafannjiang@ucdavis.edu.
The research is supported in part by The Centennial Fellowship
from American Mathematical Society,
the UC Davis Chancellor's Fellowship and
National Science Foundation grant no. DMS-0306659
}

\begin{abstract}
We consider 6 types of scaling limits
for the Wigner-Moyal equation of the parabolic waves
in random media, the limiting cases of which
include the radiative transfer limit, the diffusion limit and
the white-noise limit. We show under fairly
general assumptions on the random refractive index field
that  sufficient amount of medium diversity (thus
excluding the white-noise limit) leads to
statistical stability or self-averaging in
the sense that the limiting law is deterministic
and is governed by various transport equations
depending on the specific scaling involved. 
We obtain 6 different radiative transfer equations
as limits.
\end{abstract}

\maketitle
\section{Introduction}

The celebrated Schr\"{o}dinger equation
\[
i\hbar\frac{\partial \Psi}{\partial t} +
\frac{\hbar^2}{2m}\Delta\Psi+
{\sigma}V(t,
\bx)\Psi=0, \quad\Psi(0,\bx)=\Psi_0(\bx)
\]
describes the evolution
of the wave function $\Psi$ of a quantum spin-less particle in a potential
$\sigma V$ where ${\sigma}$ is the typical size of the variation.

A similar equation called the parabolic wave
equation is also widely used to describe the
propagation of  the modulation of low intensity
wave beam in turbulent or turbid media
in the forward scattering approximation of
the full wave equation \cite{St}. In this connection
the refractive index fluctuation plays the role
of the potential in the equation.
Nondimensionalized with respect to the 
propagation  distances in the longitudinal and transverse directions,
$L_z$ and $L_x$,
respectively,
the parabolic wave equation for the modulation function $\Psi$ reads
\beq
i\ks^{-1} L_z^{-1} \frac{\partial \Psi}{\partial z} +
2^{-1}k^{-2}L_x^{-2}\Delta\Psi+
{\sigma}V(z L_z,
{\bx L_x})\Psi=0,\quad\Psi(0,\bx)=\Psi_0(\bx)
\label{0.3}
\eeq
where $k$ is the carrier wavenumber,
$\Psi$ the amplitude modulation and
$\Delta$ the Laplacian operator
in the transverse coordinates $\bx$.
Here we have assumed the random media has a constant background.
In the sequel we will adopt the notation of
(\ref{0.3}).

\commentout{
Clear air turbulence
is an example at hand which has
multi-scale,
power-law type isotropic spectral density
\beq
\label{power}
\Phi(\bk)\sim
|\bk|^{-1-2\eta}|\bk|^{-d},\quad d=2,\quad \bk\in \IR^{d+1},
\quad \hbox{for}\,\,|\bk|\in (L_0^{-1},
\ell_0^{-1}),\quad\eta\in (0,1)
\eeq
with a slowly varying background mainly depending on
the altitude.
In particular, $\eta=1/3$ gives rise to the Kolmogorov spectrum.
In the high Reynolds number the ratio $L_0/\ell_0$ tends
to infinity.
Roughly speaking as $\ell_0\to 0$ the sample field is only
H\"{o}lder continuous with the exponent $\eta-\delta, \forall \delta>0$.
Our method and results can easily be adapted to
the case of slowing varying background.
}

In this paper we study the scaling regimes where
the wave field experiences both longitudinal and
transverse diversity of the random medium,
 represented
by
$V$,  whose
fluctuation is assumed to be weak. 
This gives rise to random spread of wave energy
in the transverse directions. 
To fix the idea, let us choose the units of the
longitudinal and transverse coordinates such that
the spectral support of the function $V$ is
$O(1)$ while
\beq
\label{0.0'}
L_z\sim L_x\gg 1,
\eeq
namely $L_z$ and $L_x$ are much larger than the 
correlation lengths of the medium in the longitudinal
and transverse directions, respectively. There
is no loss of generality in the choice of
the hyperbolic scaling (\ref{0.0'})
(cf. Remark~2 below). Depending on the actual length
scales and anisotropy of the medium,  we adjust the
intensity $\sigma$ of the medium fluctuations to obtain a
nontrivial limit. 

\section{Wigner distribution and time reversal}
There has been a surge of interest
in the radiative transfer limit in terms of the Wigner
distribution (see below) because of its application to
the spectacular phenomena related to
time-reversal (or phase-conjugate) mirrors
\cite{BPZ}, \cite{BPR2}, \cite{tire}, \cite{PRS}.

The Wigner transform or distribution of the wave
function $\Psi$ is
defined as
\beq
\nn
W(z,\bx,\bp)&=&\frac{1}{(2\pi)^d}\int e^{-i\bp\cdot\by}
\Psi (z,\bx+\frac
{\by}2){\Psi^{*}(z,\bx-\frac{\by}{2})}d\by\\
&=&\frac{1}{(2\pi)^d}\int \int e^{-i\bp\cdot(\by_1-\by_2)}
\delta(\bx-\frac{\by_1+\by_2}{2})
\rho(\by_1,\by_2)d\by_1d\by_2
\nn
\eeq
where $\rho(\by_1,\by_2)=\Psi(\by_1)\Psi^*(\by_2)$ is called the
two-point function or the density matrix.
As apparent from the definition the Wigner
function contains all the information about
$\rho$. The Wigner distribution have the simple properties
\[
\int W(z,\bx, \bp) d\bx d\bp=\|\Psi\|_2^2,\quad
\|W\|_2=(2\pi)^{-d/2}\|\Psi\|_2^2.
\]
This is the case
of pure state Wigner distribution.
What is the more pertinent for us
is the so called mixed state Wigner distribution.

Let us briefly review how a mixed state Wigner distribution
arises in the time-reversal operation.
Let $G_H(0,\bx,z,\by)$ be
the Green's function, with the point source located
at $(z,\by)$, for the reduced wave (Helmholtz) equation
for which the Schr\"odinger equation is an approximation.
By the self-adjointness of the Helmholtz equation,
$G_H$ satisfies the symmetry
property
\[
G_H(0,\bx,z,\by)=G_H(z,\by,0,\bx).
\]
The wave field $\Psi_m$ received at the mirror is given by
\bean
 \Psi_m(z, \bx_m)&=&\chi_A(\bx_m)\int
 G_H(0,\bx_m,z,\bx_s)\Psi_0(\bx_s)d\bx_s \\
 && =\chi_A(\bx_m)\int G_H(z,\bx_s,
 0,\bx_m)\Psi_0(\bx_s)d\bx_s
 \eean
 where $\chi_A$ is the aperture function of the phase-conjugating mirror $A$.

 After phase conjugation and back-propagation we have at
 the source plane the wave field
 \[
 \Psi^B(z,\bx;\ks)=\int
 G_H(z,\bx,0,\bx_m)\overline{G_H(z,\bx_s,0,\bx_m)}\chi_A(\bx_m)\overline{\Psi_0(\bx_s)}d\bx_m
 d\bx_s.
 \]
 In the parabolic approximations
 the Green's function $G_H(z,\bx,0,\by)$
 is approximated by $e^{i\ks z}G_S(z,\bx,\by)$  where
 $G_S(z,\bx,\by)$ is the propagator of the Schr\"odinger equation.
 Making the approximation in the above expression for
 the back-propagated field we obtain
 \begin{eqnarray}
 \label{eq:gpro}
 \Psi^{B}(z,\bx;\ks)
 &=&
 \int
 G_S(z, \bx, \bx_m)
 \overline{ G_S(z, \bx_s, \bx_m)
   \Psi_0\lt({\bx_s }\rt) }\chi_A(\bx_m) d\bx_m
   d\bx_s
     \nonumber  \\
	   &=& \int
	    e^{i \bp \cdot (\bx - \bx_s)}
	     W(z, \frac{\bx+\bx_s}{2},\bp)
	        \overline{\Psi_0\lt({\bx_s }\rt)}d\bp d\bx_s
		\label{eq:gpro2}
		\end{eqnarray}
		where the Wigner distribution $W$ is given by
		\bea
		\label{ww}
		\lefteqn{W(z,\bx,\bp)}\\
		 &=& \nn \frac{1}{(2\pi)^d}\int e^{-i\bp\cdot\by} G_S(z,
		 \bx+\by/2,
		 \bx_m)
		 \overline{ G_S(z,\bx-\by/2, \bx_m ) }
		 \chi_A(\bx_m)d\by d\bx_m.\nn
		 \eea
		 This is a mixed-state Wigner distribution.
		 In general, the integral in (\ref{eq:gpro2})
		 should be interpreted in the distributional sense.

The Wigner distribution in (\ref{ww}) has
the  initial condition
$W$
\bea
\label{83}
  W(0,\bx,\bp) &=& \frac{\chi_A(\bx)}{(2\pi)^d}
  \eea
  and can be treated 
  as a generalized function on $\IR^{2d}$.
  Indeed, for any $\Theta\in
  C^\infty_c(\IR^d)$ we have
  \beq
  \lan \Psi^B,\Theta\ran=
  \int \int W(z, \br, \bp) \Theta(\br, \bp)d\br d\bp
  \eeq
  where the function $\Theta$ is defined as
  \[
  \theta(\br,\bp)=2^d
  \int \Theta(\by)
  e^{i2\bp\cdot(\by-\br)/\gamma}\overline{\Psi_0(2\br-\by)}d\by.
  \]
  If for instance $\Psi_0\in C^\infty_c(\IR^d)$ then
  it is easy to see $\Theta(\by,\bp)$ is compactly supported in
  $\by\in \IR^d$ and decays rapidly (faster than any power) in $\bp\in
  \IR^d$.
   As a result we can always approximate to arbitrary accuracy the
   distributional initial data such as (\ref{83}) by square-integrable initial
   data.

The fluctuations of the back-propagated wave field
is thus determined by the fluctuations of the Wigner
distribution.
The statistical stability or self-averaging of
the Wigner distribution in turn explains, modulo the scaling
limit, the persistence and stability of the super-focusing
of the time-reversed, back-propagated wave field observed
experimentally and numerically.

Our main results show that under various scaling limits,
sufficient amount of
spatial-transverse diversity experienced
by the propagating wave pulse results in
self-averaging and 
deterministic limiting laws.

From the perspective
of the quantum stochastic dynamics
in a random environment, our results say that
due to the spatio- temporal diversity experienced
by the wave function of the quantum particle
the quantum dynamics has in the scaling limit
 a classical probabilistic description
 which is independent of the particular
 realization of the environment.
 The transition from a unitary evolution to
 an irreversible process is of course the
 outcome of the  phase-space coarse-graining by 
 the test functions. The results presented below are a
rigorous
 demonstration of decoherence, a mechanism believed
 to be responsible for the emergence of
 the classical world from the quantum one
 \cite{Zeh}, \cite{Zu}.

\section{Assumptions}
We assume $V_z(\bx)=V(z,\bx)$ is $\bx$-homogeneous
square-integrable process with
the (partial) spectral measure 
 $\widehat{V}(z,d\bq)$.
which is an orthogonal random measure
\[
\IE[\widehat{ V}(z,d\bp)\widehat{ V}(z,d\bq)]
=\delta(\bp+\bq)\Phi_0(\bp)\,\,d\bp\,d\bq
\]
and gives rise to the (partial) spectral representation of
the refractive index field
\[
V_z(\bx)\equiv V(z,\bx)=\int \exp{(i\bp\cdot \bx)}
\widehat{V}(z,d\bp).
\]
In case that $V(\xvec), \xvec\in \IR^{d+1},$ is
$\xvec$-homogeneous square-integrable random field
with the full spectral density given by $\Phi(\xi,\bk)$
we have the following relation
\[
\Phi_0(\bp)=\int \Phi(w,\bp)\,dw.
\]
    We also have the following relation between the partial and
     full spectral measures
      \[
       \hat{V}_z(d\bp)=\int e^{iz w}\hat{V}(dw,d\bp)
     \]
     such that
  \beqn
   \IE[\hat{V}_z(d\bp) \hat{V}_s(d\bq)]
    &=&\int e^{i(s-z)w}\Phi(w,\bp)\,\,dw\,\,\delta(\bp+\bq)\,\,d\bp\,d\bq\\
     &=& \check{\Phi}(s-z,\bp)\delta(\bp+\bq)\,\,d\bp\, d\bq
      \eeqn
       where
        \[
	 \check{\Phi}(s,\bp)=\int e^{isw}\Phi(w,\bp)\,\,dw.
	  \]
Since  $\Phi(\kvec)=\Phi(-\kvec),\forall\kvec\in \IR^{d+1}$, 
	   we may assume in this case that
	    \beq
	    \label{sym}
	     \Phi(w,\bq)=\Phi(-w,\bp)=\Phi(w,-\bp)=\Phi(-w,-\bp),\quad
	      \forall w, \,\,\bp
	       \eeq
	        so that $\check{\Phi}(s,\bp)$ is real-valued
		 and $\check{\Phi}(s,\bp)=\check{\Phi}(-s,\bp)$.

For simplicity of the analysis
we assume that the spectrum $\Phi$ is smooth and 
has a compact support.
We also assume the sample
field is almost surely smooth
and its spatial derivatives all have
finite moments.
The conditions of smoothness and having
a compact support are
unnecessary; they are assumed here
in order to
make simple the key estimate
(Proposition~5, 6, 7, 8).

\commentout{*******************************
By the 
properties of the
orthogonal projection  $\IE_z[\cdot]$, we know that
\beq
\label{2.48}
\IE\lt[\IE_z[\hat{V}(\triangle)]\IE_z[\hat{V}(\triangle)]\rt]
\leq \IE\lt[\hat{V}(\triangle)\hat{V}(\triangle)\rt]
=\int_\triangle \Phi(\xi,\bk)d\xi d\bk
\eeq
for every Borel set $\triangle\subset \IR^{d+1}$.

We assume 
\begin{assumption} 
The conditional spectral measure
$\IE_z\hat{V}_s(d\bk), s\geq z, $ remains a random
orthogonal measure such that
\beq
\label{orthog}
\label{power2}
\IE\lt[\IE_z[\hat{V}_s(d\bk)]
\IE_z[\hat{V}_s(d\bk')]\rt]
=\delta(\bk+\bk')
\Phi_z(\bk) d\bk d\bk'
\eeq
for some spectral density function $\Phi_z(\bk)$.
\end{assumption}

  It then follows 
from  (\ref{2.48}) that
\beq
\label{bound2}
{\Phi}_z(\bk)\leq \int\Phi(\xi,\bk) d\xi.
\eeq
***************************************}

\commentout{**********
Let $\rho(t)$ be a non-negative (random or deterministic)
function such that
\beq
\lt|\IE \lt[\IE_z[V_s(\bx)] \IE_z[V_t(\by)]\rt]\rt|
&=&\lt| \IE \lt[\IE_z[V_s(\bx)] V_t(\by)\rt]\rt|\nn\\
&\leq& \rho(s-z)\rho(t-z) \IE\lt[V_t^2\rt],\quad \forall s, t\geq
z,\forall
\bx,\by \in \IR^d.
\label{decay}
\eeq

An obvious candidate for $\rho$ is the
maximal correlation coefficient.
********}
Let $\cF_z $ and $\cF^+_z$  be the sigma-algebras generated by
$\{V_s:  \forall s\leq z\}$ and $\{V_s: \forall s\geq  z\}$,
respectively. Define
the correlation coefficient
\beq
\label{correl}
\rho(t)=\sup_{h\in \cF_z\atop \IE[h]=0,
\IE[h^2]=1}
\sup_{
g\in \cF_{z+t}^+\atop \IE[g]=0,
\IE[g^2]=1}\IE\lt[h g\rt].
\eeq

\commentout{******************
\begin{lemma}
The maximal correlation coefficient $\rho$ satisfies
the inequality (\ref{decay}). 
\end{lemma}
\begin{proof}
Let
\[
h_s(\bx)=\IE_z[V_s(\bx)],\quad g_t(\bx)= V_t(\bx).
\]
Clearly 
\beqn
h_s
&\in & L^2(P,\Omega, \cF_{z})\\
g_t &\in& \in L^2(P,\Omega,
\cF^+_{t})
\eeqn
and their second moments are uniformly bounded
in $\bx$ since
\beqn
\IE[h_s^2](\bx)&\leq& \IE[g_s^2](\bx)\\
\IE[g_s^2](\bx) &=& \int
\Phi(\xi,\bq) d\xi d\bq
\eeqn

 From the definition
(\ref{correl}) we have
\beqn
\lt|\IE[h_s(\bx) h_t(\by)]\rt|&=
\lt|\IE\lt[h_s g_t\rt]\rt|
&\leq \rho(t-z) \IE^{1/2}\lt[h_s^2(\bx)\rt]
\IE^{1/2}\lt[g_t^2\rt].
\eeqn
Hence by  setting $s=t$ first and the
Cauchy-Schwartz inequality we have
\beqn
\IE\lt[h_s^2\rt(\bx)]&\leq&\rho^2(s-z)\IE[g_t^2]\\
\IE\lt[ h_s(\bx) h_t(\by)\rt]
&\leq& \rho(t-z) \rho(s-z)
\IE[g_t^2],\quad \forall s,
t\geq z,\forall \bx, \by.
\eeqn
\end{proof}
**************}
\begin{assumption}
The correlation coefficient $\rho(t)$ is integrable
\end{assumption}

When $V_z$ is a Gaussian process, the correlation coefficient
$\rho(t)$ equals the {\em linear} correlation coefficient  $r(t)$
which has the following useful  expression
\beq
\label{corr}
r(t) &=&\sup_{g_1, g_2}
 \int \check{\Phi}(t-\tau_1-\tau_2,\bk) g_1(\tau_1,\bk)
g_2(\tau_2,\bk)d\bk d\tau_1 d\tau_2
\eeq
where the
supremum is taken over all $g_1, g_2\in L^2(\IR^{d+1})$ which are
supported on $(-\infty, 0]\times
\IR^d$ and  satisfy the constraint
\beq
\label{b.1}
\int \check{\Phi}(t-t',\bk) g_1(t,\bk)\bar{g}_1(t',\bk) dt dt' d\bk=
\int \check{\Phi}(t-t',\bk) g_2(t,\bk)\bar{g}_2(t',\bk) dt dt' d\bk=1.
\eeq
There are various criteria for the decay rate 
of the
linear correlation coefficients in the literature.
For example, according to \cite{IR}, Chapter V, Theorem~8
(after adaptation to the $t$-continuous version), even under the restrictive condition
that $\Phi\in C^\infty_c$ there is a large
class of Gaussian processes whose correlation coefficient
decays faster than any power law.

Secondly, we assume a 6-th order  quasi-Gaussian property:
Let \[
U_s^{1}(\bx)={V}_s(\bx), \quad
U^2_s(\bx)=\IE_z[{V}_s](\bx), \quad s\geq z.
\]
\begin{assumption}
For any choices of $\sigma_j\in \{1,2\}, j=1,2,...,N$ 
and a set of linear operators $\{T_j\}$, there exists
a finite constant $C$
\beq
\IE\lt[\prod_{j=1}^N
T_jU^{\sigma_j}_{s_j}(\bx_j)]\rt]&=&0,\quad N=3,5\nn\\
\lt|\IE\lt[\prod_{j=1}^NT_jU^{\sigma_j}_{s_j}(\bx_j)]\rt]\rt|
&\leq &C\sum\lt|\prod_{\widehat{mn}}
\IE\lt[T_{m}U^{\sigma_m}_{s_m}(\bx_m)
T_{n}U^{\sigma_n}_{s_n}(\bx_n)\rt]\rt|,\quad N=4,6\nn
\eeq
where the summation is over all possible pairings
$\{\widehat{mn}\}$ among $\{1,2,...,N\}$.
\end{assumption}

Finally we assume 
\begin{assumption} 
There exists a constant $C$ such that
for Theorem~1, 2, 3 (i), (iii)  and 4 (i), (iii) 
\beq 
\label{1.4}
\lim_{\ep\to
0}\IE[\sup_{z<z_0}\|\cvtil\theta\|^2_2]
\leq\frac{C}{\ep} \IE\|\cvtil\theta\|^2_2,
\quad\forall\theta\in C^\infty_c
(\IR^{2d}),\quad
\forall z_0<\infty;
\eeq
and for
Theorem~3 (ii) and ~4 (ii)
\beq
\nn
\lim_{\ep\to
0}\IE[\sup_{z<z_0}\|\cvtil\theta\|^2_2]
&\leq&\frac{C}{\ep^\al}
\IE\|\cvtil\theta\|^2_2,
\quad\forall\theta\in C^\infty_c
(\IR^{2d}),\quad
\forall z_0<\infty;\\
\label{1.4''}\nn
\lim_{\ep\to
0}\IE[\sup_{z<z_0}\|\cvtil\cvtil\theta\|^2_2]
&\leq&\frac{C}{\ep^{2\al}}
\IE\|\cvtil\cvtil \theta\|^2_2,
\quad\forall\theta\in C^\infty_c
(\IR^{2d}),\quad
\forall z_0<\infty;
\eeq
where $\cvtil$ is  defined, respectively, by
(\ref{n50}), (\ref{n51}), (\ref{n52}) and
(\ref{n53}) 
and
$\al\in (0,1)$ as specified in the
statements of the theorems.
\end{assumption}
\commentout{**************
Here $\delta_\ep V_z$ represents one of
the following expressions
\beqn
V(\frac{z}{\ep^2},\ep^{-2\alpha}\bx+\by/2)-
V(\frac{z}{\ep^2}, \ep^{-2\alpha}\bx-\by/2)& &\hbox{
for Theorem~1};\\
 \ep^{2\alpha-2}
  \lt[ V(\frac{z}{\ep^2},\frac{\bx}{\ep^{2\alpha}}-
   \by\ep^{2-2\alpha}/2)-V(\frac{z}{\ep^{2}},\frac{\bx}{\ep^{2\alpha}}+
    \by\ep^{2-2\alpha}/2)\rt] & &\hbox{ for
Theorem~2 and ~4};\\
V(\frac{z}{\ep^{2\beta}},\ep^{-2}\bx+\by/2)-
V(\frac{z}{\ep^{2\beta}}, \ep^{-2}\bx-\by/2)&
&\hbox{ for Theorem~3}.\\
\eeqn
***************}

 Assumption~3  is readily
satisfied for Gaussian random fields. This can be seen
by first observing that $\cvtil\theta$ is a Gaussian
process and $\cvtil\cvtil\theta$ is  a $\chi^2$-process
and, secondly, by an application of  Borell's 
inequality
\cite{Ad} that the supremum over $z<z_0$
inside the expectation can be over-estimated
by a 
$\log{(1/\ep)}$ factor for excursion on the
scale of any power of $1/\ep$:
\beq
\label{G1}
\IE[\sup_{z<z_0}\|\cvtil\theta\|^2_2]
&\leq&C \log{(\frac{1}{\ep})} 
\IE\|\cvtil\theta\|^2_2;\\
\label{G2}
\IE[\sup_{z<z_0}\|\cvtil\cvtil\theta\|^2_2] 
&\leq&C \log^2{(\frac{1}{\ep})} 
\IE\|\cvtil\cvtil\theta\|^2_2.
\eeq

\section{Main results}

In the standard scaling, we set
\beq
\label{0.0}
L_z=L_x=\frac{1}{\ep^2}\gg 1,\quad
\sigma=\ep.
\eeq
To describe the small scale wave energy
we consider the scaled version of the Wigner distribution
\beq
\nn
W^\eps(z,\bx,\bp)&=&\frac{1}{(2\pi)^d}\int e^{-i\bp\cdot\by}
\Psi (z,\bx+\frac
{\ep^2\by}2){\Psi^{*}(z,\bx-\frac{\epsq\by}{2})}d\by\\
&=&\frac{1}{(2\pi)^d}\int \int e^{-i\bp\cdot(\by_1-\by_2)/\ep^2}
\delta(\bx-\frac{\by_1+\by_2}{2})
\rho(\by_1,\by_2)d\by_1d\by_2.
\label{0.11}
\eeq
The Wigner distribution $W^\ep$ 
has a limit as certain
measure, the Wigner measure, introduced in \cite{LP}. 
But as remarked in the introduction, we
always consider uniformly $L^2$ initial condition
induced by
a mixed-state density matrix $\rho$.

The Wigner distribution satisfies the Wigner-Moyal equation 
\beq
\frac{\partial \wep_z}{\partial z}
+\frac{\bp}{\ks}\cdot\nabla \wep_z
+\frac{\ks}{\ep}\cv \wep_z=0
\label{wig}
\eeq
with
 $\wepz(\bx,\bp)=
 \wep(z,\bx,\bp)$.
Here 
the integral operator $\cv$ is 
given by
\beq
\label{L}
\cv\wep_z(\bx,\xtil, \bp)&=&
i\int e^{i\bq\cdot\xtil}
\lt[\wep_z(\bx,\bp+\bq/2)-\wep_z(\bx,\bp-\bq/2)\rt]
\widehat{V}(\frac{z}{\ep^2},d\bq),\quad \xtil=\bx\epal
\eeq
with $\alpha=1$. 

The more general case with $\alpha\in
(0,1)$ can be derived from a somewhat different
scaling (cf. the scaling leading to Theorem~2): We
probe a highly anisotropic medium
$V(z,\ep^{2-2\alpha}\bx)$ with the strength
\[
\sigma=\ep^{2\alpha-1}
\]
with
a wave beam composed of waves of lengths comparable
to  that of the medium, so we replace $k$ by $k
\ep^{2-2\alpha}$:
\beq
\label{k}
k\longrightarrow k \ep^{2-2\alpha}
\eeq
in the parabolic wave equation.  We then use 
the following definition of the Wigner distribution
to resolve the wave energy:
\beq
\label{new-wig}
W^\eps(z,\bx,\bp)&=&\frac{1}{(2\pi)^d}\int e^{-i\bp\cdot\by}
\Psi (z,\bx+\frac
{\ep^{2\alpha}\by}2){\Psi^{*}(z,\bx-\frac{\ep^{2\alpha}\by}{2})}d\by.
\eeq
The difference in scaling between (\ref{new-wig})
and (\ref{k}) is, of course, due to
the rescaling of coordinates (\ref{0.0}).

Since the proof of convergence is the same 
for $ \alpha\in (0,1]$ they are treated together.
 Eq. (\ref{weak2}) and its variants studied in the sequel
 are understood in the weak sense and
 we consider their weak solutions with the test function space
 $C^\infty_c(\IR^{2d})$:
 To find $W^\ep_z\in L^2([0, \infty); L^2(\IR^{2d}))$
 such that
 $\|W^\ep_z\|_2\leq \|W_0\|_2, \forall z>0,$ and
 \beq
 \lan W_z^\varepsilon, \theta\ran - \lan W_0,
 \theta \ran &=&
 \ks^{-1}\int_0^z \lan W_s^\vas, \pdgx \theta\ran ds
 +\frac{\ks}{\vas}\int_0^z \lan W_s^\vas, {\cL}^\ep_s
 \theta
 \ran ds.
 \label{weak2}
 \eeq

We shall use the notation
\[
\hat{V}^\ep_z(d\bq)=\hat{V}(\frac{z}{\ep^2},d\bq),\quad
V^\ep_z(\bx)=V(\frac{z}{\ep^2},\bx).
\]

Taking the partial inverse Fourier transform
\beq
\cF^{-1}_2\theta(\bx,\by)
\equiv\int e^{i\bp\cdot\by}\theta(\bx,\bp)\,d\bp
\nn
\eeq
we see that $\cF^{-1}_2\cv \theta (\bx,\by)$
acts
in the following completely local manner
\beq
\label{fourier}
\cF^{-1}_2\cv \theta (\bx, \xtil,\by)=
-i\delta_\ep \vep(\xtil,\by) \cF^{-1}_2 \theta(\bx,\by)
\eeq
where
\beq
\label{del}
\delta_\ep V^\ep_z(\xtil,\by)\equiv V^\ep_{z}(\xtil+\by/2)-
V^\ep_z(\xtil-\by/2).
\eeq
The operator $\cv$ is skew-symmetric and
real (i.e. mapping real-valued
functions to real-valued functions).

Since our results do not depend on the transverse dimension $d$ we hereafter
take it to be any positive integer.

\begin{remark}
Since Eq. (\ref{weak2}) is linear , the existence of weak solutions can
be established straightforwardly by  the weak-$\star$ compactness
argument. Let us briefly comment on this.
First, we introduce truncation $N<\infty$
\[
V_N(z,\bx)=V(z,\bx) ,\quad |V(z,\bx)|< N
\]
and zero otherwise.
Clearly, for such bounded $V_N$ the corresponding operator $\cv$
is a bounded self-adjoint operator on $L^2(\IR^{2d})$. Hence the
corresponding Wigner-Moyal equation preserves the $L^2$-norm
of the initial data and produces a sequence of $L^2$-bounded
weak solutions. Passing to the limit $N\to \infty$ we obtain
a $L^2$-weak solution for the original Wigner-Moyal equation
if
$V$ is locally square-integrable as is assumed here.  However,
due to the weak limiting procedure, there is no guarantee that
the $L^2$-norm of the initial data is preserved in the limit.

We will
not address the uniqueness of solution for the Wigner-Moyal
equation (\ref{weak2}) but we will show that  as $\ep\to 0$
any sequence of weak solutions to eq. (\ref{weak2}) converges
in a suitable sense to the unique solution of a deterministic
transport equation.
\end{remark}

We state our first result in the following theorem.
\begin{theorem}
\label{thm1}
Let $\Phi\in C^\infty_c$ and Assumption 1, 2, 3 be
satisfied. Then
the weak solution $W_z^\ep$ of  the Wigner-Moyal equation (\ref{weak2}), (\ref{L})
with
the initial condition $W_0\in L^2(\IR^{2d})$
converges in probability
as
the distribution-valued process 
to the deterministic limit given by the
weak solution $W_z$ of 
the radiative transfer equation
\[
\frac{\partial W_z(\bx,\bp)}{\partial z}+
\frac{\bp}{k}\cdot\nabla W_z(\bx,\bp)
=k^2\cL W_z(\bx,\bp)
\]
with the initial condition $W_0$ and one
of the following scattering operator $\cL$:
\begin{description}
\item[Case (i)] $0<\alpha<1$,
\beq
\label{rad2}
\cL W_z(\bx,\bp)
=2\pi\int\,\,\Phi(0,\bq-\bp)
[W_z(\bx,\bq)-W_z(\bx,\bp)]d\bq;
\eeq
\item[Case (ii)] $\alpha=1$, 
\beq
\label{rad}
\cL W_z(\bx,\bp)
=2\pi\int\,\, \Phi(\frac{|\bq|^2-|\bp|^2}{2k},\bq-\bp)
[W_z(\bx,\bq)-W_z(\bx,\bp)]d\bq;
\eeq
\item[Case (iii)] $\alpha>1$, 
\beq
\label{rad3}
\cL W_z
= 0.
\eeq
\end{description}
\end{theorem}
The case of $\alpha=0$ corresponds to the so-called
white-noise scaling whose limit is a Markovian
process \cite{whn-wig}. 
\commentout{
In the case
of the turbulence spectrum (\ref{power}) the scattering operators
on the right side of (\ref{rad}) and
(\ref{rad2}) are well-defined even if
$\ell_0=0$ and $L_0=\infty$ in which case
the refractive index field
is generally not Lipschitz continuous and
is not homogeneous but has homogeneous
increments.
It is possible to modify our argument
to treat the simultaneous limit
of  $\ell_0\to, L_0\to \infty$ and
$\ep\to 0$ as is done for the white-noise limit
in \cite{whn-wig}.
}
Eq. (\ref{rad}) has recently been obtained in \cite{BPR2} 
for strongly mixing $z$-Markovian refractive
index fields with  a  bounded generator. 

In order to obtain a nontrivial scattering kernel for $\alpha>1$
we need to boost up the intensity of $V$ (cf. Theorem~3).

Next we consider a second type of scaling limits
which starts with the highly anisotropic medium
$V(z,\ep^{2-2\alpha}\bx)$. We then set
 \beq
 \label{alpha}
 L_x=L_z=\ep^{-2},\quad \sigma=\ep^{2\alpha-1}, \quad 0<\alpha<1
 \eeq
  under which the parabolic wave equation becomes
   \beq
       i\ks^{-1} \frac{\partial \Psi^\ep}{\partial z} +
     2^{-1} k^{-2}\ep^2\Delta\Psi^\ep+
 \ep^{2\alpha-3} V(z\ep^{-2},
  {\bx}\ep^{-2\alpha})\Psi^\ep=0,\quad\Psi^\ep(0,\bx)=\Psi_0(\bx).
   \label{alpha2}
    \eeq
     The radiative transfer scaling (\ref{0.0}) is the limiting case
      $\alpha=1$.
       The time-evolution of the Wigner function (\ref{0.11}) is
       governed by the  Wigner-Moyal equation
 (\ref{weak2}) with
  the following operator $\cL^\ep_z$
   \beq
    \label{L2}
    \lefteqn{\cv\wep_z(\bx,\xtil,\bp)}\\
    &=&
    \nn
    i\int e^{i\bq\cdot\xtil}
     \ep^{2\alpha-2}\lt[\wep_z(\bx,\bp+\ep^{2-2\alpha}\bq/2)-\wep_z(\bx,\bp-
      \ep^{2-2\alpha}\bq/2)\rt]
       \widehat{V}^\ep_z(d\bq),\quad \xtil=\bx\epal.
 \eeq
 The partial Fourier transform of $ \cv\theta$ is now given by
 (\ref{fourier}) with the following $\delta_\ep V_z$
 \beq
 \label{del2}
 \delta_\ep V^\ep_z(\xtil,\by)=\ep^{2\alpha-2}
 \lt[ \vep(\xtil+
 \by\ep^{2-2\alpha}/2)-\vep(\xtil-
 \by\ep^{2-2\alpha}/2)\rt].
\eeq
We now state the result for the scaling limit (\ref{alpha}), (\ref{alpha2}).
\begin{theorem}
Let $0<\alpha<1$. Let $\Phi\in C^\infty_c$ and Assumption 1, 2, 3 be
satisfied. 
Then
the weak solution $W_z^\ep$ of  the Wigner-Moyal equation (\ref{weak2}), (\ref{L2})
with
the initial condition $W_0\in L^2(\IR^{2d})$
converges in probability
as
the distribution-valued process
to the deterministic limit given by the
weak solution $W_z$ of
the following advection-diffusion equations
with the initial condition $W_0$:
\beq
\label{ad}
\frac{\partial {W_z}}{\partial z}+\frac{\bp}{k}\cdot
\nabla W_z=k^2\nabla_\bp\cdot \bD\nabla_\bp W_z,
\eeq
with one of the following diffusion tensors $\bD$. 
\begin{itemize}
\item $\alpha\in (0,1)$:
\beq
\label{20}
\bD=\pi\int\Phi(0,\bq)\bq\otimes\bq d\bq;
\eeq
\item $\alpha>1$:
\[
\bD=0.
\]
\end{itemize}
\end{theorem}
For $\alpha=1$ the limit is the same as
that in Theorem~1 Case (ii). $\alpha=0$
gives rise to the white-noise limit
for the Liouville equation. 
The advection-diffusion equation (\ref{ad}) can be obtained
from (\ref{rad2}) under the diffusion limit of the latter. 

Let us consider yet another type of scaling limit
parametrized by $\beta$. We first assume a highly anisotropic
medium $V(\ep^{2-2\beta}z, \bx)$ and set
\beq
\label{beta}
L_x=L_z=\ep^{-2},\quad\sigma=\ep
\eeq
i.e. the radiative transfer scaling.
The Schr\"{o}dinger equation then becomes
   \beq
   \label{beta2}
      i\ks^{-1} \frac{\partial \Psi^\ep}{\partial z} +2^{-1}
       k^{-2}\ep^2\Delta\Psi^\ep+
 \ep^{-1} V(z\ep^{-2\beta},
  {\bx}\ep^{-2})\Psi^\ep=0,\quad\Psi^\ep(0,\bx)=\Psi_0(\bx),
  \quad \beta<1
          \eeq
	  the corresponding Wigner-Moyal equation is
\beq
\label{beta3}
\frac{\partial \wep_z}{\partial z}
+\frac{\bp}{\ks}\cdot\nabla \wep_z
+\frac{\ks}{\ep^\beta}\cv \wep_z=0
\eeq
with
\beq
\label{L3''}
\lefteqn{\cv\wep_z(\bx,\xtil,\bp)}\\
&=&
i\ep^{\beta-1}\int e^{i\bq\cdot\xtil}
\lt[\wep_z(\bx,\bp+\bq/2)-\wep_z(\bx,\bp-\bq/2)\rt]
\widehat{V}(\frac{z}{\ep^{2\beta}},d\bq),\quad\xtil=\bx\ep^{-2},\,\,
\beta<1.
\nn
\eeq

Eq. (\ref{L3''} is a borderline case of
the following family of scaling limits.
Let us consider probing
an anisotropic medium $V(\ep^{2-2\beta}z,
\ep^{2-2\alpha}
\bx), \alpha>0,$ 
with a wave beam composed of waves of lengths
comparable to that of the medium, so we switch to
(\ref{k}) and (\ref{new-wig}) for the
formulation of scaling limits. 

Two situations arise: Case (i) $\al<\beta$ and
Case (ii) $\al>\beta$. 

In the first case $\al<\beta$ 
we set 
the strength of the medium fluctuation to be
\[
\sigma=\ep^{2\al-\beta}.
\] 
The resulting equation is (\ref{beta3}) with
\beq
\label{L3'}
\cv\wep_z(\bx,\xtil,\bp)
&=&
i\int e^{i\bq\cdot\xtil}
\lt[\wep_z(\bx,\bp+\bq/2)-\wep_z(\bx,\bp-\bq/2)\rt]
\widehat{V}(\frac{z}{\ep^{2\beta}},d\bq),\quad\xtil=\bx\ep^{-2\alpha}
\eeq
In the second case $\al>\beta$ 
we set the strength
of the medium fluctuation to be 
\[
\sigma=\ep^\al.
\]
The
resulting equation is (\ref{beta3}) with
\beq
\label{L3}
\lefteqn{\cv\wep_z(\bx,\xtil,\bp)}\\
&=&
i\ep^{\beta-\alpha}\int e^{i\bq\cdot\xtil}
\lt[\wep_z(\bx,\bp+\bq/2)-\wep_z(\bx,\bp-\bq/2)\rt]
\widehat{V}(\frac{z}{\ep^{2\beta}},d\bq),\quad\xtil=\bx\ep^{-2\al}.
\nn
\eeq

We have the following theorem.
\begin{theorem}
Let $\al, \beta>0$. Let
$\Phi\in C^\infty_c$ and Assumption 1, 2, 3 be
satisfied. Then   the weak solution $W_z^\ep$ of  the
Wigner-Moyal equation
 (\ref{beta3}) with Case (i): (\ref{L3}) or Case (ii):
(\ref{L3'}) and 
the initial condition $W_0\in L^2(\IR^{2d})$
converges in probability
as
the distribution-valued process
to the deterministic limit given by the
weak solution $W_z$ of
the transport equation 
with the initial condition $W_0$:
\beq
\label{trans}
\frac{\partial {W_z}}{\partial z}+\frac{\bp}{k}\cdot
\nabla W_z=k^2\cL W_z,
\eeq
with one of the following the scattering operators
$\cL$.
\begin{description}
\item[Case (i)] $\al<\beta$
\beq
 \label{26'}
 \cL W_z(\bx,\bp)=2\pi\int 
\Phi(0,\bq-\bp)\lt[W_z(\bx,\bq)-W_z(\bx,\bp)\rt]\,d\bq.
 \eeq
\item[Case (ii)] $1<\al/\beta<4/3, d\geq 3$
\beq
 \label{26}
 \cL W_z(\bx,\bp)=2\pi\int \delta(\frac{|\bq|^2-|\bp|^2}{2k})
 \lt[\int \Phi(w,\bq-\bp)dw\rt]\lt[W_z(\bx,\bq)-W_z(\bx,\bp)\rt]\,d\bq.
 \eeq
\item[Case (iii)] $\alpha=\beta$
\beq
 \label{26''}
 \cL W_z(\bx,\bp)=2\pi\int 
\Phi(\frac{|\bq|^2-|\bp|^2}{2k},\bq-\bp)\lt[W_z(\bx,\bq)-W_z(\bx,\bp)\rt]\,d\bq.
 \eeq
\end{description}
 \end{theorem}
Theorem~3 (i) probably holds for $d=2$ and 
$\al/\beta>4/3$ but we do not pursue it here
in order to keep the argument as simple as possible.

Earlier \cite{Sp}, \cite{EY} have established
the convergence of
the mean field $\IE W^\ep_z$
for $z$-independent Gaussian
media and $d\geq 3$. Their transport equation can be
viewed as a limiting case of 
(\ref{trans}) in which $\Phi(\xi,\bk)$ is
a $\delta$-function concentrated at $\xi=0$.
See also \cite{PV} for  mean-field results 
for $z$-finitely dependent potentials.

 Unlike the transport
equations (\ref{rad}), (\ref{rad2}), the scattering
kernel (\ref{26}) is elastic in the sense that it
preserves the kinetic energy of the scattered particle
so that the incoming and outgoing momenta $\bq, \bp$
have the same magnitude.

Finally let us consider two other types of scaling limit
starting with the slowly-varying refractive index field
$V(\ep^{2-2\beta}z,\ep^{2-2\alpha}\bx),
\alpha,\beta\in (0,1)$. In the first case 
\beq
\label{case1}
\beta>\alpha,\quad 0<\alpha<1
\eeq
we set
\beq
\label{albe}
L_x=L_z=\ep^{-2},\quad \sigma=\ep^{2\alpha-\beta}
\eeq
under which we have the following parabolic wave equation
  \beq
     \label{albe2}
        i\ks^{-1} \frac{\partial \Psi^\ep}{\partial z} +
          2^{-1}k^{-2}\ep^2\Delta\Psi^\ep+
   \ep^{2\alpha-\beta-2} V(z\ep^{-2\beta},
	     {\bx}\ep^{-2\alpha})\Psi^\ep=0,\quad\Psi^\ep(0,\bx)=\Psi_0(\bx),
              \eeq
          and the corresponding Wigner-Moyal equation (\ref{beta3})
   with
   \beq
   \label{L4}
   \lefteqn{\cv\wepz(\bx,\xtil, \bp)}\\
   &=&
   i\int e^{i\bq\cdot\xtil}
   \ep^{2\alpha-2}\lt[\wep_z(\bx,\bp+\ep^{2-2\alpha}\bq/2)
   -\wep_z(\bx,\bp-\ep^{2-2\alpha}\bq/2)\rt]
   \widehat{V}(\frac{z}{\ep^{2\beta}},d\bq),\quad\xtil=\bx\epal.
   \nn
  \eeq

  In the second case 
  \beq
  \label{case2}
  \alpha>\beta, \quad 0<\alpha<1,
  \eeq we set
  \beq
  \label{albe'}
  L_x=L_z=\ep^{-2},\quad \sigma=\ep^{\alpha}.
  \eeq
  After rescaling the parabolic wave equation
  reads 
 \beq
 \label{albe2'}
 i\ks^{-1} \frac{\partial \Psi^\ep}{\partial z} +
 2^{-1}k^{-2}\ep^2\Delta\Psi^\ep+
\ep^{\alpha-2} V(z\ep^{-2\beta},
{\bx}\ep^{-2\alpha})\Psi^\ep=0,\quad\Psi^\ep(0,\bx)=\Psi_0(\bx),
\eeq
and the corresponding Wigner-Moyal equation takes the form of (\ref{beta3}) with
\beq
\label{L4'}
\lefteqn{\cv\wep_z(\bx,\xtil,\bp)}\\
&=&
i\ep^{\beta-\alpha}\int e^{i\bq\cdot\xtil}
\ep^{2\alpha-2}\lt[\wep_z(\bx,\bp+\ep^{2-2\alpha}\bq/2)
-\wep_z(\bx,\bp-\ep^{2-2\alpha}\bq/2)\rt]
\widehat{V}(\frac{z}{\ep^{2\beta}},d\bq),\xtil=\bx\epal
\nn
\eeq

\begin{theorem}
Let $\alpha, \beta\in (0,1)$. Let $\Phi\in C^\infty_c$ and Assumption 1, 2, 3 be
satisfied. Then
the weak solution $W_z^\ep$ of the Wigner-Moyal equation (\ref{beta3}) with
Case (i): (\ref{L4}) or Case (ii): (\ref{L4'}) and
the initial condition $W_0\in L^2(\IR^{2d})$
converges in probability
as
the distribution-valued process
to the deterministic limit given by the
weak solution $W_z$ of
the advection-diffusion equation (\ref{ad})
with the following diffusion tensors:
\begin{description}
\item[Case (i)-(\ref{case1}), (\ref{albe})] 
\beq
\label{20'}
\bD=\pi\int\Phi(0,\bq)\bq\otimes\bq d\bq.
\eeq
\item[Case (ii)-(\ref{case2}), (\ref{albe'})]
 $d\geq 3$, $1<\alpha/\beta<4/3$,
\beq
\label{21}
\bD(\bp)
   =\pi k|\bp|^{-1}\int
\lt[\int \Phi(w,\bp_\perp)dw\rt]\bp_\perp\otimes\bp_\perp
    d\bp_\perp  
\eeq
   where $\bp_\perp\in \IR^{d-1}, \bp_\perp\cdot\bp=0$.
   \item[Case (iii)] $\alpha=\beta$
   \beq
   \label{22}
   \bD(\bp)=\pi\int 
    \Phi(k^{-1}\bp\cdot\bq,\bq)\bq\otimes\bq\,d\bq
\eeq
\end{description}
\end{theorem}
The advection-diffusion equation with (\ref{20'}), (\ref{21})
and (\ref{22}) are the diffusion limit of
the transport equations (\ref{26}) and
(\ref{rad}), respectively.
The limiting case of $\alpha=0$ gives rise
to the white-noise model of the Liouville equation \cite{whn-wig}.
We believe that the result for Case (ii) can be
extended to $d=2$ and $\beta/\alpha\in (0,1)$.

\begin{remark}
Taken together, our results have roughly
covered all the super-parabolic scaling
\[
L_x\gg \sqrt{L_z}.
\]
To see this, let us set
\[
L_z\sim L_x^\gamma=\ep^{-2\gamma},\quad 0<\gamma<2
\]
and define the following Wigner transform
\beqn
W^\eps(z,\bx,\bp)&=&\frac{1}{(2\pi)^d}\int e^{-i\bp\cdot\by}
\Psi(z,\bx+\frac
{\tep^2\by}2){\Psi^{*}(z,\bx-\frac{\tep^2\by}{2})}d\by\\
&=&\frac{1}{(2\pi)^d}\int \int e^{-i\bp\cdot(\by_1-\by_2)/\tep^2}
\delta(\bx-\frac{\by_1+\by_2}{2})
\rho(\by_1,\by_2)d\by_1d\by_2
\eeqn
with the new parameter
\[
\tep=\ep^{2-\gamma}
\]
and analyze analogously the preceding scaling limits
as parametrized by $\tep$.
For a systematic treatment
of scaling limits resulting in a transport equation,
see \cite{rad-trans}.
\end{remark}

Our approach is to use the conditional shift \cite{Kur}
to formulate the corresponding martingale problem parametrized
by $\ep$
and adapt the perturbed test function technique
to the probabilistic setting
to establish the convergence of the
martingales. It then turns out that
after subtracting
the drift and the Stratonovich correction term
the limiting martingale
has null quadratic variation (see Proposition~\ref{quad})
implying that the
limit
is deterministic.
The perturbed test functions constructed here
(see e.g. (\ref{1st}), (\ref{2nd}) and
(\ref{3rd})) are
related to those in \cite{BPR}, \cite{BPR2} but our analysis
is carried out in a more general framework
as formulated in \cite{whn-wig} and provides a unified treatment
of a range of scaling limits from the radiative transfer
to the diffusion limit and the white-noise limit.

\section{Martingale formulation}
We consider the
weak formulation  of the Wigner-Moyal equation:
\beq
\lt[\lan W_z^\varepsilon, \theta\ran - \lan W_0,
\theta \ran\rt] &=&
\ks^{-1}\int_0^z \lan W_s^\vas, \bp\cdot\nabla \theta\ran ds
+\frac{\ks}{\vas}\int_0^z \lan W_s^\vas, \cv
\theta
\ran ds
\label{weak}
\eeq
for any test function $\theta \in C^\infty_c (\IR^{2d})$.
which is a dense subspace in $L^2(\IR^{2d}).$
The tightness result (see below) implies that
for $L^2$ initial data 
the limiting measure $\IP$ is supported
in $L^2([0,z_0]; L^2(\IR^{2d}))$.

For tightness as well as identification of the limit,
 the following infinitesimal operator  $\cA^\ep$ will play an important role.
Let $V^\vas_z\equiv V(z/\ep^2,\cdot)$. Let $\mathcal{F}_z^\vas$ be the
$\sigma$-algebras generated by $\{V_s^\vas, \, s\leq z\}$  and
$\mathbb{E}_z^\vas$ the corresponding conditional expectation w.r.t. $\cF^\ep_z$.
Let $\cM^\ep$ be the space of measurable function adapted to $\{\cF^\ep_z, \forall t\}$ 
such that $\sup_{z<z_0}\IE|f(z)|<\infty$.
We say $f(\cdot)\in \cD(\cA^\ep)$, the domain of $\cA^\ep$, and $\cA^\ep f=g$
if $f,g\in \cM^\ep$ and for
$f^\delta(z)\equiv\delta^{-1}[\IE^\ep_z f(z+\delta)-f(z)]$
we have
\beqn
\sup_{z,\delta}\IE|f^\delta(z)|&<&\infty\\
\lim_{\delta\to 0}\IE|f^\delta(z)-g(z)|&=&0,\quad\forall
z.
\eeqn
Consider a special class of admissible functions
 $f(z)=\phi(\lan W_z^\vas, \theta\ran),
 f'(z)=\phi'(\lan W_z^\vas, \theta\ran),
 \forall \phi\in C^\infty(\IR)$
we have the following expression
from (\ref{weak}) and the chain rule 
\beq
\label{gen2}
 \cA^\vas
f(z)
&=&f'(z)\lt[ \frac{1}{\ks} \lan W_z^\vas,
{\bp}\cdot \nabla\theta\ran + \frac{\ks}{\vas} \lan W_z^\vas,
\cv\theta\ran\rt].
\eeq
In case of the test function $\theta$ that is also a functional
of the media we have
\beq
\label{gen}
 \cA^\vas
 f(z)
 &=&f'(z)\lt[ \frac{1}{\ks} \lan W_z^\vas,
 {\bp}\cdot \nabla\theta\ran + \frac{\ks}{\vas} \lan W_z^\vas,
 \cv\theta\ran +\lan W_z^\ep, \cA^\ep\theta\ran\rt]
 \eeq
 and when $\theta$ depends explicitly on the fast spatial variable
 \[
 \tilde{\bx}=\bx/\ep^{2\alpha}
 \]
 the gradient $\nabla$ is conveniently decomposed into 
 the gradient w.r.t. the slow variable $\nabla_\bx$ and that
 w.r.t. the fast variable $\nabla_{\tilde{\bx}}$ 
 \[
 \nabla=\nabla_\bx+\ep^{-2\alpha}\nabla_{\tilde{\bx}}.
 \]

 A main property of $\cA^\ep$ is
that 
\beq
\label{12.2}
f(z)-\int^z_0 \cA^\ep f(s) ds\quad\hbox{is a  $\cF^\ep_z$-martingale},
\quad\forall f\in \cD(\cA^\ep).
\eeq
Also,
\beq
\label{mart}
\IE^\ep_s f(z)-f(s)=
\int^z_s \IE^\ep_s \cA^\ep f(\tau) d\tau\quad \forall s<z \quad\hbox{a.s.}
\eeq
(see \cite{Kur}).
We denote by $\cA$ the infinitesimal operator corresponding
to the unscaled process $V_z(\cdot)=V(z,\cdot)$.

\section{Proof of Theorem~1}
\subsection{Tightness}

In the sequel we will adopt the following notation 
\[
f(z)\equiv \phi(\lan  W_z^\vas,
\theta\ran),\quad
 f'(z)\equiv \phi'(\lan  W_z^\vas,
\theta\ran),\quad f''(z)\equiv
\phi''(\lan  W_z^\vas,
\theta\ran),\quad 
\quad\forall \phi\in C^\infty(\IR).
\]
Namely, the prime stands for the differentiation w.r.t. the original argument (not $t$)
of $f, f'$ etc.

Let
$D([0,\infty);L^2_{w}(\IR^{2d}))
$
be the $L^2$-valued right continuous processes with left limits
endowed with the Skorohod topology.
A family 
of processes $\{ W^\ep, 0<\ep<1\} \subset D([0,\infty);L^2_{w}(\IR^{2d})) 
$ is 
tight if and only if the family of 
processes
$\{\lan  W^\ep, \theta\ran, 0<\ep <1\}
\subset D([0,\infty);L^2_{w}(\IR^{2d}))
$
is tight for all $\theta\in C^\infty_c$ \cite{Fouque}. 
We use the tightness  criterion of \cite{Ku}
(Chap. 3, Theorem 4), namely, 
we will prove:
Firstly,
\beq
\label{trunc}
\lim_{N\to \infty}\limsup_{\ep\to 0}\IP\{\sup_{z<z_0}|\lan  W_z^\ep, \theta\ran|
\geq N\}=0,\quad\forall z_0<\infty.
\eeq
Secondly, for  each $\phi\in
C^\infty(\IR)$   there is a sequence
$f^\ep(z)\in\cD(\cA^\ep)$ such that for each $z_0<\infty$
$\{\cA^\ep f^\ep (z), 0<\ep<1,0<z<z_0\}$ 
is uniformly integrable and
\beq
\lim_{\ep\to 0} \IP\{\sup_{z<z_0} |f^\ep(z)-
\phi(\lan  W_z^\ep, \theta\ran) |\geq
\delta\}=0,\quad \forall \delta>0.
\label{n59'}
\eeq
Then it follows that the laws of
$\{\lan  W^\ep, \theta\ran, 0<\ep <1\}$ are tight in the space
of $D([0,\infty);\IR)$.

First, condition (\ref{trunc}) is satisfied because the $L^2$-norm is
uniformly bounded.

Let
\beq
\label{mix}
\lefteqn{\tilde{\cL}^\ep_z\theta(\bx,{\tilde{\bx}},\bp)}\\
&\equiv&
{i}
\ep^{-2}\int_z^\infty\int e^{i\bq\cdot{\tilde{\bx}}}
[\theta(\bx,\bp+\bq/2)-
\theta(\bx,\bp-\bq/2)] e^{i
k^{-1}(s-z)\bp\cdot\bq/\ep^{2\alpha}}
\IE_z^\ep\hat{V}^\ep_s(d\bq)
ds
\nn
\eeq
which has a compact support
since both $\theta$ and $\Phi$ do. Note
that the operator $\cvtil$ maps a real-valued
function $\theta$ to a real-valued function.

\begin{lemma}
\label{Lemma1}
\beq
\label{n5}
\IE\lt[\cltil_z\theta\rt]^2(\bx,\bp) \leq
\lt[\int^\infty_0\rho(s)ds \rt]^2
\int \lt[\theta(\bx,\bp+\bq/2)-\theta(\bx,\bp-\bq/2)\rt]^2
\Phi(\xi,\bq)d\xi d\bq
\eeq
which has an $\ep$-uniformly bounded
support and is 
bounded uniformly in $\bx,\bp,\ep$.
\end{lemma}
\begin{proof}
Consider the following trial functions
in the definition of the maximal correlation coefficient
\beq
h&=&h_{s}(\bx,\bp)\nn\\
&=&{i}
\int e^{i\bq\cdot\bx\epal} [\theta(\bx,\bp+\bq/2)-
\theta(\bx,\bp-\bq/2)]e^{i
k^{-1}(s-z)\bp\cdot\bq/\ep^{2\alpha}}
\IE_z^\ep\hat{V}^\ep_s(d\bq)\nn\\
g&=&g_{t}(\bx,\bp)\nn\\
&=&{i}
\int e^{i\bq\cdot\bx\epal} [\theta(\bx,\bp+\bq/2)-
\theta(\bx,\bp-\bq/2)]e^{i
k^{-1}(t-z)\bp\cdot\bq/\ep^{2\alpha}}
\hat{V}^\ep_t(d\bq)\nn
\eeq
It is easy to see that 
\beqn
h_s
&\in & L^2(P,\Omega, \cF_{\ep^{-2}z})\\
g_t &\in& \in L^2(P,\Omega,
\cF^+_{\ep^{-2}t})
\eeqn
and their second moments are uniformly bounded
in $\bx,\bp,\ep$ since
\beq
\IE[h_s^2](\bx,\bp)&\leq& \IE[g_s^2](\bx,\bp)\\
\IE[g_s^2](\bx,\bp) &=& \int
[\theta(\bx,\bp+\bq/2)-
\theta(\bx,\bp-\bq/2)]^2
\Phi(\xi,\bq) d\xi d\bq\label{n6}
\eeq
which is uniformly bounded for any integrable 
spectral density
$\Phi$.

 From the definition
(\ref{correl}) we have
\beqn
\lt|\IE[h_s(\bx,\bp) h_t(\by,\bq)]\rt|&=
\lt|\IE\lt[h_s g_t\rt]\rt|
&\leq \rho(\ep^{-2}(t-z) )\IE^{1/2}\lt[h_s^2(\bx,\bp)\rt]
\IE^{1/2}\lt[g_t^2(\by,\bq)\rt].
\eeqn
Hence by  setting $s=t,\bx=\by,\bp=\bq$ first and the
Cauchy-Schwartz inequality we have
\beqn
\IE\lt[h_s^2\rt(\bx,\bp)]&\leq&\rho^2(\ep^{-2}(s-z))\IE[g_t^2(\bx,\bp)]\\
\lt|\IE\lt[ h_s(\bx,\bp) h_t(\by,\bq)\rt]\rt|
&\leq& \rho(\ep^{-2}(t-z) )\rho(\ep^{-2}(s-z))
\IE^{1/2}[g_t^2(\bx,\bp)]\IE^{1/2}[g_t^2(\by,\bq)],\quad
\forall s, t\geq z,\forall \bx, \by.
\eeqn

Hence
\beqn
\ep^{-4}\int^\infty_z\int^\infty_z \IE[h_s(\bx,\bp)
g_t(\bx,\bp)] ds dt
&\leq&\IE[g_t^2](\bx,\bp)
\lt[\int^\infty_0\rho(s)ds \rt]^2
\eeqn
which together with (\ref{n6}) yields
(\ref{n5}). 
\end{proof}

\begin{cor}
\label{cor1}
\beq
\label{n9}
\lefteqn{\IE\lt[\bp\cdot\nabla_\bx\cltil_z\theta\rt]^2(\bx,\bp)}\\
& \leq&
\lt[\int^\infty_0\rho(s)ds \rt]^2
\int \lt[\bp\cdot\nabla_\bx\theta(\bx,\bp+\bq/2)-\bp\cdot\nabla_\bx\theta(\bx,\bp-\bq/2)\rt]^2
\Phi(\xi,\bq)d\xi d\bq\nn
\eeq
which has an $\ep$-uniformly bounded
support and is
bounded uniformly in $\bx,\bp,\ep$.
\end{cor}
Inequality~(\ref{n9}) can be obtained 
from the expression
\beqn
\lefteqn{\bp\cdot\nabla_{\bx}\tilde{\cL}^\ep_z\theta(\bx,{\tilde{\bx}},\bp)}\\
&\equiv&
{i}
\ep^{-2}\int_z^\infty\int e^{i\bq\cdot{\tilde{\bx}}}
\bp\cdot\nabla_\bx[\theta(\bx,\bp+\bq/2)-
\theta(\bx,\bp-\bq/2)] e^{i
k^{-1}(s-z)\bp\cdot\bq/\ep^{2\alpha}}
\IE_z^\ep\hat{V}^\ep_s(d\bq)
ds
\nn
\eeqn
as in Lemma~\ref{Lemma1}.

\commentout{*********

The partial Fourier transform of $\cvtil\theta$
has the following expression
\beq
\nn\lefteqn{\cF^{-1}_{2}\tilde{\cL}^\ep_z\theta
(\bx,\by)}\\
 &=&\frac{1}{\ep^2}\int^\infty_z
\cF^{-1}_2\theta(\bx,\by+ik^{-1}
\ep^{-2\alpha}(s-z)\nabla)
\exp{(ik^{-1}\ep^{-2\alpha} (s-z)\Delta/2)}\nn\\
&&\quad\lt[\IE^\ep_z\lt[V_s^\ep(\frac{\bx}{\ep^{2\alpha}}
-\frac{\by}{2})\rt]-\IE^\ep_z\lt[V_s^\ep(\frac{\bx}{\ep^{2\alpha}}
+\frac{\by}{2})\rt]\rt]ds\nn \\
&=& -\int \lt[
\tilde{V}^\ep_{z}(\epal{\bx}+\frac{\by}{2},\bp)-
\tilde{V}^\ep_{z}(\epal{\bx}-\frac{\by}{2},\bp)\rt]
e^{i\bp\cdot\by}
\theta(\bx,\bp)\,d\bp\label{newl}
\eeq
where
\beq
\label{new}
\tilde{V}^\ep_z(\bx,\bp)&=&\frac{1}{\ep^2}
\int^\infty_z 
e^{i\ep^{-2\alpha}k^{-1}(s-z)\Delta/2}
\IE^\ep_z\lt[ V^\ep_s(\bx-\epal k^{-1}(s-z)\bp)\rt]\,\,ds.
\nn
\eeq

Consider now the following trial functions
in the definition of the maximal correlation coefficient
\beq
h&=&h_{s}(\bx)=e^{i\ep^{-2\alpha}k^{-1}(s-z)\Delta/2}
\IE^\ep_z\lt[
V^\ep_s(\bx)\rt]
\nn\\
g&=&g_{t}(\bx)=e^{i\ep^{-2\alpha}k^{-1}(s-z)\Delta/2}
V^\ep_s(\bx).
\nn
\eeq
The same argument as that for Lemma~\ref{Lemma1}
now leads to the following.
\begin{lemma}
\label{Lemma1}
Assumption~1 implies that
\beq
\label{n7}
\IE\lt[\vtil_z\theta\rt]^2(\bx)\leq
\lt[\int^\infty_0\rho(s)ds\rt]^2 \IE\lt[V_z^2\rt]
\eeq
which is 
bounded uniformly in $\bx,\ep$.
\end{lemma}
************}

The main property of $\cltil_z\theta$ is that
it solves the corrector equation
\beq
\label{corrector}
\label{imp}
\lt[\ep^{-2\alpha}\frac{\bp}{k}\cdot\nabla_{\tilde\bx}+\cA^\ep\rt]
\tilde{\cL}^\ep_z\theta=\ep^{-2}\cL^\ep_z\theta.
\eeq
Eq. (\ref{imp}) can also be solved  by using (\ref{fourier}), yielding the solution
\beq
\label{n50}
\cF^{-1}_2\cvtil\theta(\bx,\xtil,\by)&=&
\ep^{-2}
\int_z^\infty
e^{-i\epal k^{-1}(s-z)\nabla_\by\cdot\nabla_{\tilde\bx}}
\lt[\IE^\ep_z\lt[\delta_\ep V^\ep_s\rt]
\cF^{-1}_2\theta\rt](\bx,\xtil, \by) ds
\eeq
where
\[
\delta_\ep V^\ep_z(\xtil,\by)
=V_z^\ep(\xtil+\by/2)-
V^\ep_z(\xtil-\by/2).
\]
Recall that $\grxtil$ and $\grx$ are the gradients
w.r.t. the fast variable $\xtil$ and the slow
variable $\bx$, respectively.

We will need to estimate the iteration of $\cv$
and $\cvtil$: 
\beq
\cv\cltil_z\theta(\bx,\xtil,
\bp)&=&-\ep^{-2}\int^\infty_z
\int
\hat{V}^\ep_z(d\bq)
\IE_z^\ep[\hat{V}^\ep_s(d\bq')]
e^{i\bq\cdot{\tilde{\bx}}} e^{i\bq'\cdot{\tilde{\bx}}} 
e^{i
k^{-1}(s-z)\bp\cdot\bq/\ep^{2\alpha}}\nn\\
&&\lt\{[\theta(\bx,\bp+\bq'/2+\bq/2)-
\theta(\bx,\bp+\bq'/2-\bq/2)] e^{i
k^{-1}(s-z)\bq'\cdot\bq/(2\ep^{2\alpha})}\rt.\label{n8}\\
&&\lt. -
[\theta(\bx,\bp-\bq'/2+\bq/2)-
\theta(\bx,\bp-\bq'/2-\bq/2)] e^{-i
k^{-1}(s-z)\bq'\cdot\bq/(2\ep^{2\alpha})}\rt\}
ds
\nn\\
\cltil_z\cltil_z\theta(\bx,\xtil,
\bp)&=&-\ep^{-4}\int_z^\infty\int^\infty_z
\int
\IE_z^\ep[\hat{V}^\ep_s(d\bq)]
\IE_z^\ep[\hat{V}^\ep_t(d\bq')]
e^{i\bq\cdot{\tilde{\bx}}} e^{i\bq'\cdot{\tilde{\bx}}} 
e^{i
k^{-1}(s-z)\bp\cdot\bq/\ep^{2\alpha}}e^{i
k^{-1}(t-z)\bp\cdot\bq'/\ep^{2\alpha}}\nn\\
&&\lt\{[\theta(\bx,\bp+\bq'/2+\bq/2)-
\theta(\bx,\bp+\bq'/2-\bq/2)] e^{i
k^{-1}(s-z)\bq'\cdot\bq/(2\ep^{2\alpha})}\rt.\label{n8'}\\
&&\lt. -
[\theta(\bx,\bp-\bq'/2+\bq/2)-
\theta(\bx,\bp-\bq'/2-\bq/2)] e^{-i
k^{-1}(s-z)\bq'\cdot\bq/(2\ep^{2\alpha})}\rt\}
ds dt.
\nn
\eeq
Both have a compact support. Their second moments
can be estimated as in Lemma~\ref{Lemma1}
by using the 6-th order quasi-Gaussian property
(Assumption~2). In order to carry out the same argument
we need to approximate
the terms of non-product form such as
$\theta(\bx,\bp+\bq'/2+\bq/2) e^{i
k^{-1}(s-z)\bq'\cdot\bq/(2\ep^{2\alpha})}$
by sum of product of functions of
variables that are statistically coupled 
in the quasi-Gaussian pairing. 

Since we do not need the pointwise estimate
such as stated in Lemma~\ref{Lemma1} we shall
demonstrate a simpler approach based on the inverse
Fourier transform:
\beq
\cF^{-1}_2\lt\{\cv\cltil_z\theta\rt\}(\bx,\xtil,
\by)\label{n7'} &=&
\ep^{-2}\int^\infty_z
\delta_\ep
V^\ep_z e^{-i\epal
k^{-1}(s-z)\nabla_\by\cdot\nabla_{\tilde
\bx}}\lt[\IE_z[\delta_\ep V^\ep_s]\cF^{-1}_2\theta\rt]
(\bx,\xtil, \by)ds\\
\label{prob-est}
\cF^{-1}_2\lt\{\cltil_z\cltil_z\theta\rt\}(\bx,\xtil,
\by)\label{n7''} &=&
-\ep^{-4}\int^\infty_z
e^{-i\epal k^{-1}(t-z)\nabla_\by\cdot\nabla_{\tilde
\bx}}\lt\{\IE_z[\delta_\ep
V^\ep_t]e^{-i\epal k^{-1}(s-z)
\nabla_\by\cdot\nabla_{\tilde
\bx}}\rt.\\
&&\lt. \hspace{5cm} \lt[\IE_z[\delta_\ep
V^\ep_s]\cF^{-1}_2\theta\rt]\rt\} (\bx,\xtil, \by)ds
dt.\nn
\eeq
\begin{lemma}
\label{Lemma2}
\beq
\IE\|\cv\cvtil\theta\|_2^2
&\leq &8C\lt(\int^\infty_0\rho(s)ds\rt)^2
 \IE[ V_z]^2
  \int 
    [\theta(\bx,\bp+\bq/2)-
       \theta(\bx,\bp-\bq/2)]^2\Phi(\xi, \bq)d\xi d\bx d\bq d\bp
       \nn\\
\IE\|\cvtil\cvtil\theta\|_2^2
&\leq &8C\lt(\int^\infty_0\rho(s)ds\rt)^4
 \IE[ V_z]^2
  \int 
    [\theta(\bx,\bp+\bq/2)-
       \theta(\bx,\bp-\bq/2)]^2\Phi(\xi, \bq)d\xi d\bx d\bq d\bp
       \nn
       \eeq
       for some constant $C$ independent of $\ep$.
\end{lemma}
\begin{proof}
The calculation for  $\cv\cvtil\theta$ is simpler.
Let us consider $\cvtil\cvtil\theta$.

By the Parseval theorem and the unitarity of
$\exp{(i\tau \gry\cdot \grxtil)},\tau\in \IR,$
\beqn
\lefteqn{\IE\|\cltil_z\cltil_z\theta\|_2^2=
\IE\|\cF^{-1}_2\cltil_z\cltil_z\theta\|_2^2}\\
&=&\ep^{-8}\int\IE\lt\{\int^\infty_z
e^{-i\epal k^{-1}(t-z)\nabla_\by\cdot\nabla_{\tilde
\bx}}\lt\{\IE_z[\delta_\ep
V^\ep_t]e^{-i\epal k^{-1}(s-z)\nabla_\by\cdot\nabla_{\tilde
\bx}}\lt[\IE_z[\delta_\ep V^\ep_s]\cF^{-1}_2\theta\rt]\rt\}
(\bx,\xtil, \by)ds dt\rt.\\
&&\lt.\int^\infty_z
e^{i\epal k^{-1}(t'-z)\nabla_\by\cdot\nabla_{\tilde
\bx}}\lt\{\IE_z[\delta_\ep
V^\ep_{t'}]e^{i\epal
k^{-1}(s'-z)\nabla_\by\cdot\nabla_{\tilde
\bx}}\lt[\IE_z[\delta_\ep V^\ep_{s'}]\cF^{-1}_2\theta\rt]\rt\}
(\bx,\xtil,\by)ds'
dt'\rt\}d\bx
d\by\\&=&
\ep^{-8}\int\IE\lt\{\int^\infty_z
e^{-i\epal k^{-1}(t-t')/2\nabla_\by\cdot\nabla_{\tilde
\bx}}\lt\{\IE_z[\delta_\ep
V^\ep_t]e^{-i\epal k^{-1}(s-z)\nabla_\by\cdot\nabla_{\tilde
\bx}}\lt[\IE_z[\delta_\ep V^\ep_s]\cF^{-1}_2\theta\rt]\rt\}
(\bx,\xtil,\by)ds dt\rt.\\
&&\lt.\int^\infty_z
e^{i\epal k^{-1}(t-t')/2\nabla_\by\cdot\nabla_{\tilde
\bx}}\lt\{\IE_z[\delta_\ep
V^\ep_{t'}]e^{i\epal
k^{-1}(s'-z)\nabla_\by\cdot\nabla_{\tilde
\bx}}\lt[\IE_z[\delta_\ep V^\ep_{s'}]\cF^{-1}_2\theta\rt]\rt\}
(\bx,\xtil,\by)ds' dt'\rt\}d\bx d\by\\
&=&\ep^{-8}\int\IE\lt\{\int^\infty_z
\lt\{\IE_z[\delta_\ep
V^\ep_t]e^{-i\epal k^{-1}(s-z)\nabla_\by\cdot\nabla_{\tilde
\bx}}\lt[\IE_z[\delta_\ep V^\ep_s]\cF^{-1}_2\theta\rt]\rt\}
(\bx,\xtil,\by)ds dt\rt.\\
&&\lt.\int^\infty_z
\lt\{\IE_z[\delta_\ep
V^\ep_{t'}]e^{i\epal
k^{-1}(s'-z)\nabla_\by\cdot\nabla_{\tilde
\bx}}\lt[\IE_z[\delta_\ep V^\ep_{s'}]\cF^{-1}_2\theta\rt]\rt\}
(\bx,\xtil, \by)ds' dt'\rt\}d\bx d\by\\
&\leq&C\ep^{-8}\int\int^\infty_z
\lt|\IE\lt\{\IE_z[\delta_\ep
V^\ep_t]e^{-i\epal k^{-1}(s-z)\nabla_\by\cdot\nabla_{\tilde
\bx}}\lt[\IE_z[\delta_\ep V^\ep_s]\cF^{-1}_2\theta\rt]\rt\}
(\bx,\xtil, \by)\rt|ds dt\\
&&\int^\infty_z
\lt|\IE\lt\{\IE_z[\delta_\ep
V^\ep_{t'}]e^{i\epal
k^{-1}(s'-z)\nabla_\by\cdot\nabla_{\tilde
\bx}}\lt[\IE_z[\delta_\ep V^\ep_{s'}]\cF^{-1}_2\theta\rt]\rt\}
(\bx,\xtil,\by)\rt|ds' dt'd\bx d\by\\
&&+C\ep^{-8}\int\int^\infty_z
\lt|\IE\lt[\IE_z[\delta_\ep
V^\ep_t]\IE_z[\delta_\ep
V^\ep_{t'}]\rt]\rt| \lt|\IE\lt\{e^{-i\epal
k^{-1}(s-z)\nabla_\by\cdot\nabla_{\tilde
\bx}}\lt[\IE_z[\delta_\ep V^\ep_s]\cF^{-1}_2\theta\rt]
(\bx,\by)\rt.\rt.\\
&&\lt.\lt.
\quad\quad \times e^{i\epal
k^{-1}(s'-z)\nabla_\by\cdot\nabla_{\tilde
\bx}}\lt[\IE_z[\delta_\ep V^\ep_{s'}]\cF^{-1}_2\theta\rt]
(\bx,\xtil,\by)\rt\}\rt|ds dt ds' dt'd\bx d\by.
\eeqn
The last inequality follows from the quasi-Gaussian
property. Note that in the $\bx$ integrals above the
fast variable
$\xtil$ is integrated and not treated 
as independent of $\bx$.

Let
\[
g(t)=\delta_\ep
V^\ep_t
\]
and
\[
h(s)=e^{-i\epal k^{-1}(s-z)\nabla_\by\cdot\nabla_{\tilde
\bx}}\lt[\delta_\ep V^\ep_s\cF^{-1}_2\theta\rt].
\]
The same argument as that for Lemma~\ref{Lemma1} yields
\beqn
\lt|\IE[\IE_z[g(t)]\IE_z[ h(s)]]\rt|&\leq&
\IE^{1/2}[\IE_z[g(t)]^2]\IE^{1/2}[\IE_z[h(s)]^2]\\
&\leq&\rho(\ep^{-2}(t-z))\rho(\ep^{-2}(s-z))
\IE^{1/2}[g^2(t)]\IE^{1/2}[h^2(s)],\quad t, s\geq z;\\
\lt|\IE[\IE_z[g(t)]\IE_z[ g(t')]]\rt|&\leq&
\IE^{1/2}[\IE_z[g(t)]^2]\IE^{1/2}[\IE_z[g(t')]^2]\\
&\leq&\rho(\ep^{-2}(t-z))\rho(\ep^{-2}(t'-z))
\IE^{1/2}[g^2(t)]\IE^{1/2}[g^2(t')],\quad t, t'\geq z;\\
\lt|\IE[\IE_z[h(s)]\IE_z[ h(s')]]\rt|&\leq&
\IE^{1/2}[\IE_z[h(s)]^2]\IE^{1/2}[\IE_z[h(s')]^2]\\
&\leq&\rho(\ep^{-2}(s-z))\rho(\ep^{-2}(s'-z))
\IE^{1/2}[h^2(s)]\IE^{1/2}[h^2(s')],\quad s, s'\geq
z.
\eeqn

Combining the above estimates we get
\beqn
\IE\|\cltil_z\cltil_z\theta\|_2^2&\leq&
2C\lt(\int^\infty_0\rho(s)ds\rt)^4
\int\IE[g(z)]^2\IE[h(z)]^2d\bx d\by
\\
&\leq& 
2C\lt(\int^\infty_0\rho(s)ds\rt)^4
\int \IE[\delta_\ep V^\ep_z]^2
\IE\lt[e^{-i\epal k^{-1}(s-z)\nabla_\by\cdot\nabla_{\tilde
\bx}}\lt[\delta_\ep V^\ep_s\cF^{-1}_2\theta\rt]\rt]^2d\bx d\by\\
&\leq&8C\lt(\int^\infty_0\rho(s)ds\rt)^4
\IE[ V^\ep_z]^2
\int\IE\lt[e^{-i\epal k^{-1}(s-z)\nabla_\by\cdot\nabla_{\tilde
\bx}}\lt[\delta_\ep V^\ep_s\cF^{-1}_2\theta\rt]\rt]^2d\bx d\by\\
&\leq&8C\lt(\int^\infty_0\rho(s)ds\rt)^4
\IE[ V^\ep_z]^2
\int \IE
 \lt\{e^{i\bq\cdot{\tilde{\bx}}}
 [\theta(\bx,\bp+\bq/2)-
 \theta(\bx,\bp-\bq/2)]\rt.\nn\\
 &&\lt. \hspace{6cm} \times
 e^{i
 k^{-1}(s-z)\bp\cdot\bq/\ep^{2\alpha}}
 \hat{V}^\ep_s(d\bq)\rt\}^2 d\bx d\bp\\
 &\leq&8C\lt(\int^\infty_0\rho(s)ds\rt)^4
 \IE[ V^\ep_z]^2
 \int 
  [\theta(\bx,\bp+\bq/2)-
   \theta(\bx,\bp-\bq/2)]^2\Phi(\xi, \bq)d\xi d\bx d\bq d\bp
\eeqn

\end{proof}

Eq. (\ref{n7''}) is convenient for estimating
the second moment of 
$\bp\cdot\nabla_\bx\cvtil\cvtil\theta$
and $\cv\cvtil\cvtil\theta$ which by (\ref{n7''}) and
(\ref{fourier}) have the following expressions
\beq
\lefteqn{\cF^{-1}_2\lt\{\pdgx\cltil_z\cltil_z
\theta\rt\}(\bx,\xtil,\by)}\label{n10}\nn\\
&=&i
\ep^{-2}\nabla_\by\cdot\nabla_\bx \int^\infty_z
e^{-i\epal k^{-1}(t-z)\nabla_\by\cdot\nabla_{\tilde
\bx}}\lt\{\IE_z[\delta_\ep
V^\ep_t]e^{-i\epal k^{-1}(s-z)\nabla_\by\cdot\nabla_{\tilde
\bx}}\lt[\IE_z[\delta_\ep V^\ep_s]\cF^{-1}_2\theta\rt]\rt\}
(\bx,\by)ds dt\nn\\
&=&i
\ep^{-2} \int^\infty_z
e^{-i\epal k^{-1}(t-z)\nabla_\by\cdot\nabla_{\tilde
\bx}}\lt\{\IE_z[\delta_\ep
V^\ep_t]e^{-i\epal k^{-1}(s-z)\nabla_\by\cdot\nabla_{\tilde
\bx}}\lt[\IE_z[\nabla_\by\delta_\ep
V^\ep_s]\cdot\cF^{-1}_2\nabla_\bx\theta\rt]\rt\} (\bx,\by)ds
dt\nn\\ &&+
i\ep^{-2} \int^\infty_z
e^{-i\epal k^{-1}(t-z)\nabla_\by\cdot\nabla_{\tilde
\bx}}\lt\{\IE_z[\nabla_\by\delta_\ep
V^\ep_t]\cdot e^{-i\epal
k^{-1}(s-z)\nabla_\by\cdot\nabla_{\tilde
\bx}}\lt[\IE_z[\delta_\ep
V^\ep_s]\cF^{-1}_2\nabla_\bx\theta\rt]\rt\} (\bx,\by)ds
dt\nn
\eeq

\beq
\lefteqn{\cF^{-1}_2\lt\{\cv\cltil_z\cltil_z\theta\rt\}(\bx,\xtil,\by)}\label{n10'}\nn\\
&=&i
\ep^{-4} \delta_\ep V^\ep_z(\xtil,\by)\int^\infty_z
e^{-i\epal k^{-1}(t-z)\nabla_\by\cdot\nabla_{\tilde
\bx}}\lt\{\IE_z[\delta_\ep
V^\ep_t]e^{-i\epal k^{-1}(s-z)\nabla_\by\cdot\nabla_{\tilde
\bx}}\lt[\IE_z[\delta_\ep V^\ep_s]\cF^{-1}_2\theta\rt]\rt\}
(\bx,\by)ds dt.\nn
\eeq
The same calculation as in Lemma~\ref{Lemma2} yields 
the following estimates:
\begin{cor}
\label{cor2}
\beqn
\lefteqn{\IE\|\bp\cdot\nabla_\bx\cltil_z\cltil_z\theta\|_2^2}\\
&\leq&
32C\lt(\int^\infty_0\rho(s)ds\rt)^4\lt\{
 \IE[\nabla_\by V^\ep_z]^2
 \int 
  [\nabla_\bx\theta(\bx,\bp+\bq/2)-
   \nabla_\bx\theta(\bx,\bp-\bq/2)]^2\Phi(\xi, \bq)d\xi d\bx
d\bq d\bp\rt.\\
&&\hspace{3cm}+\lt.
 \IE[V^\ep_z]^2
 \int 
  [\nabla_\bx\theta(\bx,\bp+\bq/2)-
   \nabla_\bx\theta(\bx,\bp-\bq/2)]^2|\bp|^2\Phi(\xi,
\bq)d\xi d\bx d\bq d\bp\rt\};
\eeqn
\beqn
\IE\|\cv\cltil_z\cltil_z\theta\|_2^2
&\leq&
32C\lt(\int^\infty_0\rho(s)ds\rt)^4
 \IE[ V^\ep_z]^4
 \int 
  [\theta(\bx,\bp+\bq/2)-
   \theta(\bx,\bp-\bq/2)]^2\Phi(\xi, \bq)d\xi d\bx d\bq d\bp
\eeqn
for some constant $C$ independent of $\ep$.
\end{cor}

Let
\begin{equation}
\label{1st}
f_1^\vas (z)=
{\ks\vas}
f'(z) \lan  W_z^\vas,\cvtil\theta\ran
\end{equation}
be the 1-st perturbation of $f(z)$. 

\begin{prop}\label{prop:2}
\begin{enumerate}
$$\lim_{\ep\to 0}\sup_{z<z_0} \mathbb{E} |f_1^\vas(z)|=0,\quad
\lim_{\ep\to 0}\sup_{z<z_0} |f_1^\vas(z)|= 0
\quad \hbox{in probability}$$.
\end{enumerate}
\end{prop}

\begin{proof}
We have
\beq
\label{1.2}
\mathbb{E}[|f_1^\vas(z)|]\leq {\vas}\|f'\|_\infty
\| W_0\|_2 
\IE\|\tilde{\cV}^\ep_z\theta\|_2
\eeq
and
\beq
\label{1.3}
\sup_{z< z_0} |f_1^\vas(z)|
 \leq {\vas}
\|f'\|_\infty  \| W_0\|_2
\sup_{z<z_0}\|\cvtil\theta\|_2.
\eeq
The right side of (\ref{1.2}) is $O(\ep)$ while
the right side of (\ref{1.3}) is $o(1)$ in probability
by Chebyshev's inequality and Assumption~3.

Proposition~\ref{prop:2}
now follows from (\ref{1.2}) and (\ref{1.3}).
\end{proof}

Set $f^\ep(z)=f(z)+f^\ep_1(z)$.
A straightforward calculation yields
\beqn
\cA^\vas f_1^\vas &=&
{\ep} f'(z)\lan\wepz,\bp\cdot\nabla_\bx\cvtil \theta\ran
+\ep f''(z)
\lan \wepz, \bp\cdot\nabla_\bx
\theta\ran\lan\wepz, \cltil\theta\ran\\
&&+{k^2}f'(z)\lan
\wepz,\cv\cvtil\theta\ran+
{\ks}^2 f''(z)\lan \wepz, \cL_z^\ep \theta\ran\lan \wepz, 
    \cvtil\theta\ran\\
 &&
 -\frac{\ks}{\ep}f'(z)\lan \wepz,\cv\theta\ran
\nn
\eeqn
and, hence
\beq
\label{25'}
\cA^\vas f^\ep(z)
&=&\frac{1}{\ks}f'(z)\lan \wepz, \pdgx \theta\ran
+{\ks^2}f'(z)\lan \wepz, \cv\cvtil\theta\ran
+{\ks^2}f''(z)\lan\wepz,\cv\theta\ran\lan\wepz,\cvtil\theta\ran
\nn\\
&&\quad
+{\ep}\lt[f'(z)\lan\wepz,\pdgx\cvtil\theta\ran
+f''(z)\lan\wepz,\pdgx\theta\ran\lan \wepz,\cvtil\theta\ran\rt]\\
&=&A_0^\vas(z)+A_1^\vas(z)+A_2^\vas(z)+R_1^\vas(z)
\nonumber
\eeq
where $A_1^\vas(z)$ and $A_2^\vas(z)$ are the $O(1)$ statistical coupling
terms.

\begin{prop}
\label{prop:3}
$$\lim_{\ep\to
0}\sup_{z<z_0}\IE|R_1^\ep(z)|^2=0$$.
\end{prop}
\begin{proof}
 \beq
 |R^\ep_1|
  \nonumber
  &\leq& {\vas}
  \left[\|f''\|_\infty\| W_0\|^2_2
  \|\pdgx\theta\|_2\|\cvtil\theta\|_2+
  \|f'\|_\infty\|\psiep\|_2\|\pdgx(\cvtil\theta)\|_2\rt].
 \label{1.10}
  \eeq
Clearly $$\lim_{\ep\to
0}\sup_{z<z_0}\IE|R^\ep_1(z)|^2=0.$$
 by
Lemma~\ref{Lemma1}
  and Corollary~\ref{cor1}.  
\end{proof}

For the tightness criterion stated in the beginnings of the section,
it remains to show
\begin{prop}
\label{prop:1}
$\{\cA^\ep f^\ep\}$ are uniformly integrable.
\end{prop}

\begin{proof}
We show that $\{A^\ep_i\}, i=0, 1,2,3$ are uniformly integrable.

For this we have the following estimates:
\bean
|A_0^\vas(z)|
&\leq&\frac{1}{\ks}\|f'\|_\infty\| W_0\|_2\|\pdgx \theta\|_2\\
|A_1^\vas(z)|
&\leq& {\ks^2}\|f'\|_\infty\| W_0\|_2
\|\cv\cvtil\theta\|_2\\
|A_2^\vas(z)|
&\leq& {\ks^2}\|f''\|_\infty\| W_0\|^2_2
\|\cv\theta\|_2\|\cvtil\theta\|_2.
\eeqn
The second moments of the right hand side
of the above expressions are uniformly bounded as
$\ep\to 0$ by Lemmas~\ref{Lemma1} and \ref{Lemma2}
and hence $A_0^\vas(z), A_1^\vas(z), A_2^\vas(z)$ are uniformly
integrable. 
By Proposition~\ref{prop:3}, $R^\ep_1$
is uniformly integrable.
  \end{proof}

\subsection{Identification of the limit}
Our strategy  is to show directly
that in passing to the weak limit  the limiting process
solves the martingale problem with zero quadratic variation.
The uniqueness
of the  limiting deterministic problem  then identifies the limit.

For this purpose,
we introduce the next perturbations $f_2^\ep, f_3^\ep$.
Let
\bea
\label{40.3}
A_2^{(1)}(\psi) &\equiv&\int\psi(\bx,\bp)
\cQ_1(\theta\otimes\theta)(\bx,\bp,\by,\bq)\psi(\by,\bq)\,d\bx d\bp\,d\by d\bq\\
A_1^{(1)}(\psi)&\equiv&\int\cQ'_1\theta(\bx,\bp)\psi(\bx,\bp)\,\,d\bx d\bp,\quad
\forall \psi\in L^2(\IR^{2d})
\label{41.2}
\eea
where 
\beq
    \cQ_1(\theta\otimes\theta)(\bx,\bp,\by,\bq)&=&
    \IE\lt[\cv\theta(\bx,\bp)\cvtil\theta(\by,\bq)\rt]
\eeq
and
\[
\cQ'_1\theta(\bx,\bp)=\IE\lt[\cv\cvtil\theta (\bx,\bp)\rt].
\]
Clearly,
\beq
A_2^{(1)}(\psi)=\mathbb{E}\lt[\lan\psi, \cv\theta\ran
\lan\psi, \cvtil\theta\ran\rt].
\eeq

Let
\[
\cQ_2(\theta\otimes\theta)(\bx,\bp,\by,\bq)\equiv
\IE\lt[\cvtil\theta(\bx,\bp)\cvtil\theta(\by,\bq)\rt]
\]
and \[
\cQ'_2\theta(\bx,\bp)=\IE\lt[\cvtil\cvtil\theta (\bx,\bp)\rt].
\]
Let
\bea
A_2^{(2)}(\psi) &\equiv&\int\psi(\bx,\bp)
\cQ_2(\theta\otimes\theta)(\bx,\bp,\by,\bq)\psi(\by,\bq)\,\,d\bx d\bp\,d\by
d\bq\\
A_1^{(2)}(\psi)&\equiv&\int\cQ_2'\theta(\bx,\bp)\psi(\bx,\bp)\,\,d\bx\,d\bp
\eea
Define
\beq
\label{2nd}
f_2^\vas(z) &=&\frac{\ep^2\ks^2}{2}f''(z)
\lt[\lan\psiep,\cvtil\theta\ran^2-A^{(2)}_2(\psiep)\rt]\\
f_3^\vas(z) &=&
\frac{\ep^2\ks^2}{2}f'(z)\lt[\lan\psiep,\cvtil\cvtil\theta\ran-
A^{(2)}_1(\psiep)\rt].
\label{3rd}
\eeq

\begin{prop}\label{prop:4}
$$ \lim_{\ep\to 0}\sup_{z<z_0} \mathbb{E}|f_2^\vas(z)|=0,\quad \lim_{\ep\to 0}\sup_{z<z_0}
\mathbb{E}|f_3^\vas(z)|=0. $$
\end{prop}
\begin{proof}
We have the bounds
\beqn
\sup_{z<z_0}\IE|f_2^\vas(z)|&\leq&
\sup_{z<z_0}{\vas^2\ks^2}\|f''\|_\infty\lt[
\| W_0\|_2^2\IE\|\cvtil\theta\|_2^2
+\IE[A^{(2)}_2(\psiep)]\rt]\\
\sup_{z<z_0}
\IE|f_3^\vas(z)|&\leq&
\sup_{z<z_0}
{\vas^2\ks^2}\|f'\|_\infty
\lt[\| W_0\|_2\IE \|\cvtil\cvtil\theta\|_2
+\IE[A_1^{(2)}(\psiep)]\rt].
\eeqn
The right sides of both tend to zero as $\ep\to 0$
by Lemma~1 and 2.
\end{proof}

We have
\beqn
\cA^\vas f_2^\vas(z)&=&{\ks^2}
f''(z)\left[-
\lan\tvas, \cv\theta\ran\lan\tvas,\cvtil\theta\ran + A^{(1)}_2(\tvas)\right]
+ R_2^\vas(z)\\
\cA^\vas f_3^\vas(z)&=&{\ks^2}
f'(z)\left[-\lan
\tvas,\cv(\cvtil\theta)\ran + A^{(1)}_1(\tvas)\right]+
R_3^\vas(z)
\eeqn
with
\beq
\nn
R_2^\vas(z)& =&\ep^2{\ks^2}\frac{f'''(z)}{2}\left[
\frac{1}{\ks}
\lan\psiep,\pdgx\theta\ran+\frac{\ks}{\ep}\lan\psiep,\cv\theta\ran\rt]
\lt[\lan\psiep,\cvtil\theta\ran^2-A_2^{(2)}(\psiep)\rt]\\
\nn
&& +
\ep^2{\ks^2}f''(z)\lan\psiep,\cvtil\theta\ran\lt[
\frac{1}{\ks}\lan\psiep,\pdgx(\cvtil\theta)\ran+\frac{\ks}{\ep}
\lan\psiep,\cv\cvtil\theta\ran\rt]\\
&&-\ep^2{\ks^2}f''(z)
\lt[\frac{1}{\ks}\lan\psiep,\pdgx(G_\theta^{(2)}\psiep)\ran+
\frac{\ks}{\ep}\lan\psiep,\cv G_\theta^{(2)}\psiep\ran\rt]
\label{36}
\eeq
where 
$G_\theta^{(2)}$ denotes the operator
\[
G_\theta^{(2)}\psi\equiv \int \cQ_2(\theta\otimes\theta)(\bx,\bp,\by,\bq)\psi(\by,\bq)
\,d\by d\bq.
\]
Similarly
\beq
\label{n75}
R^\ep_3(z)&=&\ep^2{\ks^2}
f'(z)\left[\frac{1}{\ks}\lan\psiep,\pdgx(\cvtil\cvtil\theta)\ran+\frac{\ks}{\ep}
\lan\psiep,\cv\cvtil\cvtil\theta\ran\rt]\\
&&\nn +\ep^2\frac{\ks^2}{2}
f''(z)\left[
\frac{1}{\ks}\lan\psiep,\pdgx\theta\ran+\frac{\ks}{\ep}\lan\psiep,\cv\theta\ran\rt]
\lt[\lan\psiep,\cvtil\cvtil\theta\ran-A_1^{(2)}(\psiep)\rt]\\
&&\nn-\ep^2{\ks^2}f'(z)\lt[
\frac{1}{\ks}\lan\psiep,\pdgx(\cQ_2'\theta)\ran+\frac{\ks}{\ep}
\lan\psiep,\cv\cQ_2'\theta\ran\rt].
\eeq

\begin{prop}\label{prop:5}
\[
\lim_{\ep\to 0}\sup_{z<z_0} \mathbb{E} |R_2^\vas(z)|=0,\quad \lim_{\ep \to 0}
\sup_{z<z_0} \mathbb{E}
|R_3^\vas(z)|=0.
\]
\end{prop}
\begin{proof}
Part of the argument is analogous to that  given for
Proposition~\ref{prop:4}.  The additional estimates that we need to consider
are  the following. 

In $R_2^\ep$: First
\beqn
\label{r2.1}\nn
\lefteqn{\sup_{z<z_0}\ep^2
\IE\lt|\lan\psiep,\pdgx(G_\theta^{(2)}\psiep)\ran\rt|}\\ 
&=&\ep^2\int \IE\lt[ \psiep(\bx,\bp)\psiep(\by,\bq)\rt]\IE\lt[
\bp\cdot\nabla_\bx
\cvtil\theta(\bx,\bp)\cvtil\theta(\by,\bq)\rt] 
d\bx d\by d\bp d\bq\\
&\leq& \ep^2\int \IE\lt[ \psiep(\bx,\bp)\psiep(\by,\bq)\rt]
\IE^{1/2}[\bp\cdot\nabla_\bx \cvtil\theta]^2(\bx,\bp)\IE^{1/2}[
\cvtil \theta]^2(\by,\bq) d\bx d\by d\bp d\bq
\commentout{********************
&=&
\ep^{-2}\IE\lt\{\int\int_z^\infty\int_z^\infty \int 
e^{i\bp'\cdot{\tilde{\bx}}}e^{i\bq'\cdot{\tilde{\by}}}
 e^{i
k^{-1}(s-z)\bp\cdot\bp'/\ep^{2\alpha}}
e^{i
k^{-1}(t-z)\bq\cdot\bq'/\ep^{2\alpha}} [\theta(\by,\bq+\bq'/2)-
\theta(\by,\bq-\bq'/2)]\nn\rt.\\
&&\lt.
\times\bp\cdot\nabla_\bx [\theta(\bx,\bp+\bp'/2)-
\theta(\bx,\bp-\bp'/2)]\IE\lt[ \psiep(\bx,\bp)\psiep(\by,\bq)\rt]
\IE_z^\ep[\hat{V}^\ep_s(d\bp')] \hat{V}^\ep_t(d\bq')
ds dt d\bx d\by d\bp d\bq\rt\}
********************}
\nn
\eeqn
which is $O(\ep^2)$ by using Lemma~\ref{Lemma1},
Corollary~\ref{cor1} and the fact  $ \IE\lt[
\psiep(\bx,\bp)\psiep(\by,\bq)\rt]\in L^2(\IR^{4d})$ 
in conjunction with the same
argument as in proof of Lemma~1; Secondly,
\beqn
\label{r2.2}
\sup_{z<z_0}\ep \IE\lt|\lan\psiep,\cv G_\theta^{(2)}
\psiep\ran\rt|
&=&\sup_{z<z_0}\ep\|W_0\|_2 \IE\|\cv
G_\theta^{(2)}\psiep\|_2\nn\\
&=&\sup_{z<z_0}\ep\|W_0\|_2 \IE\|\cv
\IE\lt[\cvtil\theta\otimes\cvtil\theta\rt]\psiep\|_2\nn\\
&=&\sup_{z<z_0}\ep\|W_0\|_2 \IE\|\invf\cv
\IE\lt[\invf\cvtil\theta\otimes\invf\cvtil\theta\rt]\invf\psiep\|_2.\nn
\commentout{***********************
&=&\sup_{z<z_0}\ep\|W_0\|_2
\IE\|\ep^{-4} \delta_\ep V^\ep_z(\bx,\by)\int^\infty_z
e^{-i\epal k^{-1}(t-z)\nabla_\by\cdot\nabla_{\tilde
\bx}}\IE\lt\{\IE_z[\delta_\ep
V^\ep_t]e^{-i\epal k^{-1}(s-z)\nabla_\by\cdot\nabla_{\tilde
\bx}}\lt[\IE_z[V^\ep_s]\cF^{-1}_2\theta\rt]\rt\}
(\bx,\by)ds dt\|_2\nn\\
&\leq&\ep\|W_0\|_2 \IE\lt|\ep^{-4}\int^\infty_z\int^\infty_z 
\int
 e^{i\bp'\cdot\bx} e^{ik^{-1}(s-z)\bp\cdot\bp'\epal}
e^{\bp''\cdot\bx \epal}
\ep^{2\alpha-2}
 \lt[\theta(\bx,\bp+\bp'/2+\ep^{2-2\alpha}\bp''/2)\rt.\rt.\\
&&\lt.\lt.
-\theta(\bx,\bp-\bp'/2+\ep^{2-2\alpha}\bp''/2)
-\theta(\bx,\bp+\bp'/2-\ep^{2-2\alpha}\bp''/2)
+\theta
(\bx,\bp-\bp'/2-\ep^{2-2\alpha}\bp''/2)\rt]\hat{V}^\ep_z(d\bp'')
e^{i\bq'\cdot\by}\rt.\\
&&\lt. e^{ik^{-1}(t-z)\bq\cdot\bq'\epal} 
[\theta(\by,\bq+\bq'/2)-\theta(\by,\bq-\bq'/2)]\IE_z^\ep[\hat{V}^\ep_s(d\bp')]
\IE_z[\hat{V}^\ep_t(d\bq')] d\bx d\bp ds dt \psiep(\by,\bq) d\by
d\bq\nn
\rt|\\
&\leq& \ep\|W_0\|^2_2
\IE^{1/2}\lt\|\ep^{-4}\int^\infty_z\int^\infty_z 
\int
 e^{i\bp'\cdot\bx} e^{ik^{-1}(s-z)\bp\cdot\bp'\epal}
e^{\bp''\cdot\bx \epal}
\ep^{2\alpha-2}
 \lt[\theta(\bx,\bp+\bp'/2+\ep^{2-2\alpha}\bp''/2)\rt.\rt.\\
&&\lt.\lt.
-\theta(\bx,\bp-\bp'/2+\ep^{2-2\alpha}\bp''/2)
-\theta(\bx,\bp+\bp'/2-\ep^{2-2\alpha}\bp''/2)
+\theta
(\bx,\bp-\bp'/2-\ep^{2-2\alpha}\bp''/2)\rt]\hat{V}^\ep_z(d\bp'')
e^{i\bq'\cdot\by}\rt.\\
&&\lt. e^{ik^{-1}(t-z)\bq\cdot\bq'\epal} 
[\theta(\by,\bq+\bq'/2)-\theta(\by,\bq-\bq'/2)]
\IE\lt[\IE_z^\ep[\hat{V}^\ep_s(d\bp')]
\hat{V}^\ep_t(d\bq')\rt]
d\bx d\bp ds dt\rt\|_2^2 \nn
***********************}
\eeqn
Let
\beqn
h_s=e^{-ik^{-1}\epal(s-z)\nabla_\by\cdot\nabla_{\tilde\bx}} [\delta_\ep V^\ep_z \invf\theta].
\eeqn
We then have
\beqn
\label{n11}
\lefteqn{\IE\|\invf\cv
\IE\lt[\invf\cvtil\theta\otimes\invf\cvtil\theta\rt]\invf\psiep\|_2
}\\
&=&\IE\lt\{\int
\lt|\ep^{-4}\int \int^\infty_z \delta_\ep V^\ep_z(\bx,\by)
\IE\lt[\IE_z[ h_s(\bx,\by)]\IE_z[h_t(d\bx',d\by')]\rt]
\invf \wepz(\bx', \by')d\bx' d\by' ds dt\rt|^2 d\bx d\by\rt\}^{1/2}\\
&\leq&\IE^{1/2}\lt\{\int
\lt|\ep^{-4}\int^\infty_z |\delta_\ep V^\ep_z(\bx,\by)|
\rho(\ep^{-2}(s-z))\rho(\ep^{-2}(t-z))
\IE^{1/2}[h_s(\bx,\by)]^2\rt. \rt.\\
&&\lt.\lt.\quad
\int \IE^{1/2}[h_t(d\bx',d\by')]^2
|\invf \wepz(\bx', \by')|d\bx' d\by' ds dt\rt|^2 d\bx d\by\rt\}^{2}\\
&\leq&
\IE^{1/2}\lt\{\int
\lt|\ep^{-4}\int^\infty_z |\delta_\ep V^\ep_z(\bx,\by)|
\rho(\ep^{-2}(s-z))\rho(\ep^{-2}(t-z))
\IE^{1/2}[h_s(\bx,\by)]^2\rt. \rt.\\
&&\lt.\lt.\quad
\lt(\int \IE[h_t(d\bx',d\by')]^2 d\bx'd\by'\rt)
\lt(\int |\wepz(\bx', \bp')|^2d\bx' d\bp'\rt) ds dt\rt|^2 d\bx d\by\rt\}.
\eeqn
Recall that $\|W^\ep_z\|_2\leq \|W_0\|_2$ and
\beqn
\int \IE[h_t(d\bx',d\by')]^2 d\bx'd\by'
=\int
[\theta(\bx,\bp+\bq/2)-
\theta(\bx,\bp-\bq/2)]^2
\Phi(\xi,\bq) d\xi d\bq d\bx d\bp<\infty
\eeqn
so that
\beqn
\lefteqn{\IE\|\invf\cv
\IE\lt[\invf\cvtil\theta\otimes\invf\cvtil\theta\rt]\invf\psiep\|_2
}\\
&\leq& \|W_0\|_2 \IE^{1/2}\|h_s\|_2^2 \IE^{1/2}\lt\{\int
\lt|\ep^{-4}\int^\infty_z |\delta_\ep V^\ep_z(\bx,\by)|
\rho(\ep^{-2}(s-z))\rho(\ep^{-2}(t-z))
\IE^{1/2}[h_s(\bx,\by)]^2
ds dt\rt|^2 d\bx d\by\rt\}\\
&\leq& \|W_0\|_2 \IE^{1/2}\|h_s\|_2^2\lt( \sup_{\bx,\by}{
\IE[\delta_\ep V^\ep_z]^2}\rt)
\ep^{-8}\int^\infty_z 
\rho(\ep^{-2}(s-z))\rho(\ep^{-2}(t-z))
\\
&&\hspace{5cm}\times\rho(\ep^{-2}(s'-z))\rho(\ep^{-2}(t'-z))
\IE^{1/2}\|h_s\|_2^2
\IE^{1/2}\|h_{s'}\|_2^2
ds dt ds' dt' \\
&\leq& \|W_0\|_2 \IE^{3/2}\|h_s\|_2^2\lt( \sup_{\bx,\by}{
\IE[\delta_\ep V^\ep_z]^2}\rt)
\lt|\int^\infty_{0} 
\rho(s)
ds \rt|^2<\infty.
\eeqn
Recall from (\ref{n6}) that
\beqn
\IE\|h_s\|_2^2&=&\int
[\theta(\bx,\bp+\bq/2)-
\theta(\bx,\bp-\bq/2)]^2
\Phi(\xi,\bq) d\xi d\bq d\bx d\bp <\infty.\label{n16}
\eeqn
Hence 
\[
\sup_{z<z_0}\ep \IE\lt|\lan\psiep,\cv G_\theta^{(2)}\wepz\ran\rt|=O(\ep). 
\]

In $R_3^\ep$:
\beqn
\label{r3.1}
{\sup_{z<z_0}\ep 
\IE\lt|\lan\psiep,\cv\cvtil\cvtil\theta\ran\rt|}
 &\leq& \ep
\|W_0\|_2\sup_{z<z_0}\IE\|\cv\cvtil\cvtil\theta\|_2
\eeqn
which is $O(\ep)$ by Corollary~\ref{cor2}.

\commentout{*************************************
&\leq&\ep
\|W_0\|_2\sup_{z<z_0}\IE^{1/2}\lt\|\ep^{-4}\int_z^\infty\int^\infty_z
\int e^{i\bq\cdot\tilde{\bx}} \hat{V}_z(d\bp')
\IE_z^\ep\hat{V}^\ep_s(d\bq) \IE_z^\ep\hat{V}^\ep_t(d\bq')
e^{i\bq\cdot{\tilde{\bx}}}
e^{i\bq'\cdot{\tilde{\bx}}} 
e^{i
k^{-1}(s-z)\bp\cdot\bq/\ep^{2\alpha}}e^{i
k^{-1}(t-z)\bp\cdot\bq'/\ep^{2\alpha}}\rt.\nn\\
&&\lt.\ep^{2\alpha-2}\lt\{e^{i
k^{-1}(s-z)\bp'\cdot\bq\ep^{2-4\alpha}}e^{i
k^{-1}(t-z)\bp'\cdot\bq'\ep^{2-4\alpha}}\lt[[\theta(\bx,\bp+\bp'/2+\bq'/2+\bq/2)-
\theta(\bx,\bp+\bp'/2+\bq'/2-\bq/2)]\rt. \rt.\rt.\nn\\
&&\lt.\lt.\lt.\times e^{i
k^{-1}(s-z)\bq'\cdot\bq/(2\ep^{2\alpha})} -
[\theta(\bx,\bp+\ep^{2-2\alpha}\bp'/2-\bq'/2+\bq/2)-
\theta(\bx,\bp+\ep^{2-2\alpha}\bp'/2-\bq'/2-\bq/2)]\rt.\rt.\rt.\\
&&\lt.\lt.\lt.\times e^{-i
k^{-1}(s-z)\bq'\cdot\bq/(2\ep^{2\alpha})}\rt]-e^{-i
k^{-1}(s-z)\bp'\cdot\bq\ep^{2-4\alpha}}e^{-i
k^{-1}(t-z)\bp'\cdot\bq'\ep^{2-4\alpha}}\lt[[\theta(\bx,\bp-
\ep^{2-2\alpha}\bp'/2+\bq'/2+\bq/2)\rt.\rt.\rt.\\
&&\lt.\lt.\lt.-
\theta(\bx,\bp-\ep^{2-2\alpha}\bp'/2+\bq'/2-\bq/2)]  e^{i
k^{-1}(s-z)\bq'\cdot\bq/(2\ep^{2\alpha})} 
-[\theta(\bx,\bp-\ep^{2-2\alpha}\bp'/2-\bq'/2+\bq/2)-\rt.\rt.\rt.\nn\\
&&\lt.\lt.\lt.-
\theta(\bx,\bp-\ep^{2-2\alpha}\bp'/2-\bq'/2-\bq/2)] e^{-i
k^{-1}(s-z)\bq'\cdot\bq/(2\ep^{2\alpha})}\rt]\rt\}
ds dt\rt\|^2_2
\eeqn
which is $O(\ep)$ by using Assumption~2.
**********************************}
The other two terms in
$R^\ep_3$
\beqn
\label{r3.2}
\ep^2
\IE\lt|\lan\psiep,\pdgx(\cQ_2'\theta)\ran\rt|&\leq&
\ep^2\|W_0\|_2\IE^{1/2}\|\pdgx
\IE[\cvtil\cvtil\theta]\|_2^2\\
&\leq& \ep^2\|W_0\|_2\|\IE[\pdgx\cvtil\cvtil\theta]\|_2
\eeqn
which is $O(\ep^2)$ by Corollary~\ref{cor2};
\beqn
{\ep}
\IE\lt|\lan\psiep,\cv\cQ_2'\theta\ran\rt|&\leq&
\ep\|W_0\|_2\IE\|\cv\IE[\cvtil\cvtil\theta]\|_2\\
&\leq&\ep\|W_0\|_2\IE\|\invf[\cv]\IE[\invf\cvtil\cvtil\theta]\|_2\\
&\leq&\ep\|W_0\|_2\lt(\sup_{\bx,\by}\IE^{1/2}\lt|\delta_\ep
V^\ep_z\rt|^2\rt) \IE^{1/2}\|\cvtil\cvtil\theta\|_2^2
\eeqn
which is $O(\ep)$ by Lemma~\ref{Lemma2}.

\end{proof}

Consider the test function
$f^\ep(z)=f(z)+f_1^\ep(z)
+f_2^\ep(z)+f_3^\ep(z)$. We have
\beq
\label{2.67}
\lefteqn{\cA^\ep f^\ep(z)}\\
&=&
\frac{1}{\ks}f'(z)\lan\psiep,\pdgx\theta\ran+{\ks^2}f''(z)
A_2^{(1)}(\psiep)+{\ks^2}f'A_1^{(1)}(\psiep)
+R_1^\ep(z)+R_2^\ep(z)+R_3^\ep(z).\nn
\eeq
Set
\beq
\label{remainder}
R^\vas(z) = R_1^\vas(z) + R_2^\vas(z) +
R_3^\vas(z).
\eeq
It follows from Propositions 3 and 5 that
\[
\lim_{\ep \to 0}\sup_{z<z_0}\IE|R^\ep(z)|=0.
\]

\begin{prop}
\label{quad}
\[
\lim_{\ep\to 0}\sup_{z<z_0}\sup_{\|\psi\|_2=1}
A_2^{(1)}(\psi)
=0.
\]
\end{prop}
\begin{proof}
We have
\beqn
A_2^{(1)}(\psi) &=&\int\psi(\bx,\bp)
\cQ_1(\theta\otimes\theta)
(\bx,\bp,\by,\bq)\psi(\by,\bq)\,d\bx d\bp\,d\by d\bq\\
&=&\frac{1}{2}\int\psi(\bx,\bp)
\lt[\cQ_1(\theta\otimes\theta)
(\by,\bq,\bx,\bp)+\cQ_1(\theta\otimes\theta)
(\bx,\bp,\by,\bq)
\rt]
\psi(\by,\bq)\,d\bx d\bp\,d\by d\bq
\eeqn
where
the kernel $\cQ_1(\theta\otimes\theta)$
can be written as
\beqn
\lefteqn{\cQ_1(\theta\otimes\theta)
(\by,\bq,\bx,\bp)+\cQ_1(\theta\otimes\theta)
(\bx,\bp,\by,\bq)
}\\
&=&
\int^\infty_{-\infty}ds
\int \,d\bp' 
\check{\Phi}(s,\bp')
e^{i\bp'\cdot(\bx-\by)/\ep^{2\alpha}}e^{-ik^{-1} s\bp\cdot\bp'\ep^{2-2\alpha}}
\lt[\theta(\bx,\bp+\bp'/2)-\theta(\bx,\bp-\bp'/2)\rt]\\
&&\hspace{8cm}\times\lt[\theta(\by,\bq+\bp'/2)-\theta(\by,\bq-\bp'/2)\rt]\\
&=&
2\pi\int
e^{i\bp'\cdot(\bx-\by)/\ep^{2\alpha}}\lt[\theta(\bx,\bp+\bp'/2)-\theta(\bx,\bp-\bp'/2)\rt]
\\
&&\times\lt[\theta(\by,\bq+\bp'/2)-\theta(\by,\bq-\bp'/2)\rt]
\Phi(k^{-1}\bp\cdot\bp'\ep^{2-2\alpha}, \bp')d\bp'
\eeqn
which is uniformly compactly supported on $\IR^{4d}$.
For smooth and compactly supported $\Phi$, $\cQ_1(\theta\otimes\theta)$
tends to zero fast than any power of $\ep$ uniformly outside
any neighborhood of $\bx=\by$ while stays uniformly
bounded everywhere. Therefore the $L^2$-norm
of $\cQ_1(\theta\otimes\theta)$ tends to zero
and the proposition follows.
\end{proof}

Similar calculation leads to the following
expression: For any real-valued, $L^2$-weakly
convergent sequence $\psi^\ep \to\psi$, we have
\beqn
\lefteqn{\lim_{\ep\to 0}A_1^{(1)}(\psi)}\\
&=&
\lim_{\ep\to 0} \int^\infty_0ds\int dw 
d\bq d\bx d\bp \,\,\psi^\ep(\bx,\bp)
\Phi(w,\bq) e^{isw}e^{-ik^{-1}s\bp\cdot\bq \ep^{2-2\alpha}}\\
&&\quad\lt[e^{-ik^{-1}s|\bq|^2\ep^{2-2\alpha}/2}
\lt[\theta(\bx,\bp+\bq)-\theta(\bx,\bp)\rt]
-e^{ik^{-1}s|\bq|^2\ep^{2-2\alpha}/2}
\lt[\theta(\bx,\bp)-\theta(\bx,\bp-\bq)\rt]
\rt]\\
&=&\lim_{\ep\to0}
\int^\infty_0ds \int
d\bq d\bx d\bp \,\,\psi^\ep(\bx,\bp)
\check{\Phi}(s,\bq)
e^{-ik^{-1}s\bp\cdot\bq
\ep^{2-2\alpha}}\\
&&\quad\lt[e^{-ik^{-1}s|\bq|^2\ep^{2-2\alpha}/2}
\lt[\theta(\bx,\bp+\bq)-\theta(\bx,\bp)\rt]
-e^{ik^{-1}s|\bq|^2\ep^{2-2\alpha}/2}
\lt[\theta(\bx,\bp)-\theta(\bx,\bp-\bq)\rt]\rt].
\eeqn
Note that the integrand is invariant under the
change of variables:
\[
s\to -s,\quad \bq\to -\bq.
\]
Thus we can write 
\beqn
\lefteqn{\lim_{\ep\to 0}A_1^{(1)}(\psi)}\\
&=&\lim_{\ep\to0}
\frac{1}{2}\int^\infty_{-\infty}ds \int
d\bq d\bx d\bp \,\,\psi^\ep(\bx,\bp)
\check{\Phi}(s,\bq)
e^{-ik^{-1}s\bp\cdot\bq
\ep^{2-2\alpha}}\\
&&\quad\lt[e^{-ik^{-1}s|\bq|^2\ep^{2-2\alpha}/2}
\lt[\theta(\bx,\bp+\bq)-\theta(\bx,\bp)\rt]
-e^{ik^{-1}s|\bq|^2\ep^{2-2\alpha}/2}
\lt[\theta(\bx,\bp)-\theta(\bx,\bp-\bq)\rt]\rt]\\
&=&
\lim_{\ep\to0}\pi\int d\bq d\bx d\bp 
\,\,\psi^\ep(\bx,\bp)
\lt\{\Phi(\ep^{2-2\alpha}k^{-1}(\bp+\bq/2)\cdot\bq,\bq)
\lt[\theta(\bx,\bp+\bq)-\theta(\bx,\bp)\rt]\right.\\
&&\left.\quad
-\Phi(\ep^{2-2\alpha}k^{-1}(\bp-\bq/2)\cdot\bq,\bq)
\lt[\theta(\bx,\bp)-\theta(\bx,\bp-\bq)\rt]\rt\}\\
&=&
\lim_{\ep\to0}
\pi\int d\bq d\bx d\bp \,\,\psi^\ep(\bx,\bp)
\lt\{\Phi(\ep^{2-2\alpha}\frac{|\bq|^2-|\bp|^2}{2 k},
\bq-\bp)\lt[\theta(\bx,\bq)-
\theta(\bx,\bp)\rt]\rt.\\
&&\lt.\hspace{5cm} -
\Phi(\ep^{2-2\alpha}\frac{|\bp|^2-|\bq|^2}{2 k}, \bp-\bq)
\lt[\theta(\bx,\bp)-
\theta(\bx,\bq)\rt]\rt\}\\
&=&\lim_{\ep\to0}
2\pi\int d\bq d\bx d\bp \,\,\psi^\ep(\bx,\bp)
\Phi(\ep^{2-2\alpha}\frac{|\bp|^2-|\bq|^2}{2 k}, \bp-\bq)
\lt[\theta(\bx,\bq)-\theta(\bx,\bp)\rt]
\eeqn
Clearly, we have that
\beqn
\lefteqn{\lim_{\ep\to 0}\int
d\bq [\theta(\bx,\bq)-\theta(\bx,\bp)]
\Phi(\ep^{2-2\alpha}\frac{|\bq|^2-|\bp|^2}{2 k}, \bq-\bp)
}\\
&=&\int
d\bq [\theta(\bx,\bq)-\theta(\bx,\bp)]\times\left\{\begin{array}{ll}
\Phi((2k)^{-1}(|\bq|^2-|\bp|^2),\bq-\bp),&\alpha=1\\
\Phi(0,\bq-\bp),& 0<\alpha<1\\
0,&\alpha>1
\end{array}
\right.
\eeqn
in $L^2(\IE^{2d})$. Therefore,
\beqn
\lim_{\ep\to 0} {A}^{(1)}_1(\psi^\ep)
&=&\int
d\bq d\bx d\bp \psi(\bx,\bp)
[\theta(\bx,\bq)-\theta(\bx,\bp)]\times\left\{\begin{array}{ll}
\Phi((2k)^{-1}(|\bq|^2-|\bp|^2),\bq-\bp),&\alpha=1\\
\Phi(0,\bq-\bp),& 0<\alpha<1\\
0,&\alpha>1
\end{array}
\right.\\
&\equiv&\bar{A}_1(\psi)
\eeqn

Recall that
\beq
\label{64}
M_z^\vas(\theta)&=&f^\ep(z)-\int^z_0 \cA^\ep f^\ep(s)\,ds\\
&=&
f(z)+f_1^\vas(z)+f_2^\vas(z)+f_3^\vas(z)
-
\int_0^z\frac{1}{\ks}f'(z)\lan\tvas,\pdgx\theta\ran\,ds\nn\\
&& - \int_0^z{\ks^2}\left[f''(s)
A_2^{(1)}( W_s^\vas)+f'(s) A_1^{(1)}( W_s^\vas)\right]\,ds -\int_0^z
R^\vas(s)\,ds\nn
\eeq
is a martingale.
The martingale property implies that for any finite
sequence
$ 0<z_1<z_2<z_3<...<z_n \leq z $,  $C^2$-function $f$ and 
bounded continuous  function $h$ with compact support,
we have 
\beq
\label{78}
\IE\lt\{h \lt(\lan W^\ep_{z_1}, \theta\ran,
\lan W^\ep_{z_2},\theta\ran,...,\lan W^\ep_{z_n},\theta\ran\rt)
\lt[M^\ep_{z+s}(\theta)-M^\ep_{z}(\theta)\rt]\rt\}&=&0,\\
\quad \forall
s>0,\quad  z_1\leq z_2\leq\cdots\leq z_n\leq z.&&\nn
\eeq
Let
\[
\bar{\cA} f(z)\equiv
f'(s)\left[\frac{1}{\ks}\lan
 W_z,\pdgx\theta\ran+{\ks^2}\bar{A}_1( W_z)\right].
\]

In view of the results of
Propositions~\ref{prop:2},\ref{prop:3}, \ref{prop:1},
\ref{prop:4}, 
\ref{prop:5}  we see
that
$f^\ep(z)$ and
$\cA^\ep f^\ep(z)$ in (\ref{64}) can be
replaced by $f(z)$ and
$\bar{\cA}f(z)$, respectively, modulo an error
that vanishes as
$\ep\to 0$.
With this and the tightness of $\{W^\ep_z\}$ 
we can pass to the limit
$\ep\to 0$ in  (\ref{78}).  
We see that the limiting process satisfies the 
martingale property that
\beqn
\IE\lt\{h \lt(\lan W_{z_1}, \theta\ran,
\lan W_{z_2},\theta\ran,...,\lan W_{z_n},\theta\ran\rt)
\lt[M_{z+s}(\theta)-M_{z}(\theta)\rt]\rt\}=0,\quad \forall s>0.
\eeqn
where
\begin{equation}
M_z(\theta)=f(z)-\int_0^z \bar{\cA}f(s)
\,ds.
 \label{37'}
\end{equation}
Then it follows that
\[
\IE\lt[ M_{z+s}(\theta)-M_z(\theta)|W_u, u\leq z\rt]=0,\quad
\forall z,s>0
\]
which proves that $M_z(\theta)$ is a martingale given by
\begin{equation}
M_z(\theta)=f(z)-\int_0^z
\bigg\{f'(s)\left[\frac{1}{\ks}\lan
 W_s,\pdgx\theta\ran+{\ks^2}\bar{A}_1( W_s)\right]
 \bigg\}\,ds.
 \label{37}
\end{equation}
Choose $f(r) =r$ and $r^2$ in (\ref{37})
we see that
\[
M_z^{(1)}(\theta)=\lan  W_z,\theta\ran -
\int_0^z \left[\frac{1}{\ks}\lan  W_s,\pdgx\theta\ran +
{\ks^2}\bar{A}_1( W_s)\right]\,ds \]
is a martingale with the
null quadratic variation
\[ 
\left[M^{(1)}(\theta),M^{(1)}(\theta)\right]_z=
0.
\]
Thus
\[
f(z)-\int_0^z
\bigg\{f'(s)\left[\frac{1}{\ks}\lan
 W_s,\pdgx\theta\ran+{\ks^2}\bar{A}_1( W_s)\right]
  \bigg\}\,ds= f(0),\quad\forall z>0.
  \]

Since $\lan\tvas,\theta\ran$ is uniformly bounded
\[
\lt|\lan\tvas,\theta\ran\rt|\leq \| W_0\|_2{\|\theta\|}_2
\]
we have the convergence of the second moment
\[
\lim_{\ep\to 0}
\IE\left\{
{\lan\tvas,\theta\ran}^2\right\}={\lan
 W_z,\theta\ran}^2
 \]
 and hence the convergence in probability.

\section{Proof of Theorem~2}

\subsection{Tightness}
Instead of (\ref{mix}) we use the following the
corrector
\beq
\label{mix2}
\lefteqn{\tilde{\cL}^\ep_z\theta(\bx,\tilde{\bx},\bp)}\\
&=&
\frac{i}{\ep^2}
\int_z^\infty
\int e^{i\bq\cdot\xtil}e^{i
k^{-1}(s-z)\bp\cdot\bq/\ep^{2\alpha}}
\ep^{2\alpha-2}[\theta(\bx,\bp+\ep^{2-2\alpha}\bq/2)-
\theta(\bx,\bp-\ep^{2-2\alpha}\bq/2)]\IE_z^\ep\hat{V}^\ep_s(d\bq)
\nn
\eeq
which satisfies the corrector equation (\ref{corrector}).
Its inverse Fourier transform is given by
\beq
\cF^{-1}_2\cvtil\theta(\bx,\xtil,\by)&=&
\ep^{-2}
\int_z^\infty
e^{-i\epal k^{-1}(s-z)\nabla_\by\cdot\nabla_{\tilde\bx}}
\lt[\IE_z\lt[\delta_\ep V^\ep_s\rt]
\cF^{-1}_2\theta\rt](\bx,\by) ds.
\label{n51}
\eeq
Instead of Lemma~\ref{Lemma1}, Corollary~\ref{cor1}, Lemma~\ref{Lemma2}
and Corollary~\ref{cor2} we have
\begin{lemma}
\label{Lemma2.1}
Assumption~1 implies that
\beq
\label{n13}
\limsup_{\ep\to 0}\IE\lt[\cltil_z\theta\rt]^2(\bx,\bp) \leq
\lt[\int^\infty_0\rho(s)ds \rt]^2
\int\lt[\bq\cdot \nabla_\bp\theta(\bx,\bp)\rt]^2
\Phi(\xi,\bq)d\xi d\bq
\eeq
which has a bounded
support and is
bounded uniformly in $\bx,\bp$.
\end{lemma}

\begin{cor}
\label{cor2.1}
\beq
\label{n15}
\limsup_{\ep\to 0}
\IE\lt[\bp\cdot\nabla_\bx\cltil_z\theta\rt]^2(\bx,\bp)
& \leq&
\lt[\int^\infty_0\rho(s)ds \rt]^2
\int \lt[\bp\cdot\nabla_\bx[\bq\cdot\nabla_{\bp}\theta(\bx,\bp)]\rt]^2
\Phi(\xi,\bq)d\xi d\bq
\nn
\eeq
which has a bounded
support and is
bounded uniformly in $\bx,\bp$.
\end{cor}

\begin{lemma}
\label{Lemma2.2}
\beq
\limsup_{\ep\to 0}
\IE\|\cv\cvtil\theta\|_2^2
&\leq &8C\lt(\int^\infty_0\rho(s)ds\rt)^2
 \IE[ V_z]^2
  \int 
    [\bq\cdot\nabla_\bp\theta(\bx,\bp)]^2\Phi(\xi,
\bq)d\xi d\bx d\bq d\bp
       \nn\\
\limsup_{\ep\to 0}\IE\|\cltil\cltil\theta\|_2^2
&\leq &8C\lt(\int^\infty_0\rho(s)ds\rt)^4
 \IE[ V_z]^2
  \int
    [\bq\cdot\nabla_\bp\theta(\bx,\bp)
       ]^2\Phi(\xi, \bq)d\xi d\bx d\bq d\bp
       \nn
       \eeq
       for some constant $C$.
\end{lemma}
\begin{cor}
\label{cor2.2}
\beqn
\limsup_{\ep\to 0}
\IE\|\cv\cltil_z\cltil_z\theta\|_2^2
&\leq&
32C\lt(\int^\infty_0\rho(s)ds\rt)^4
 \IE[ V_z]^4
 \int
  [\bq\cdot\nabla_\bp\theta(\bx,\bp)
   ]^2\Phi(\xi, \bq)d\xi d\bx d\bq d\bp;\\
\limsup_{\ep\to
0}\IE\|\bp\cdot\nabla_\bx\cltil_z\cltil_z\theta\|_2^2 &\leq&
32C\lt(\int^\infty_0\rho(s)ds\rt)^4\lt\{
 \IE[\nabla_\by V_z]^2
 \int 
  [\bq\cdot\nabla_\bp\nabla_\bx\theta(\bx,\bp)]^2\Phi(\xi,
\bq)d\xi d\bx d\bq d\bp\rt.\\
&&\hspace{3cm}+\lt.
 \IE[V_z]^2
 \int 
[\grx\bq\cdot\nabla_\bp\theta(\bx,\bp)]^2|\bp|^2\Phi(\xi,
\bq)d\xi d\bx d\bq d\bp\rt\}
\eeqn
for some constant $C$.
\end{cor}

\commentout{*********
We have that
\beq
\cF^{-1}_{2}\tilde{\cL}^\ep_z\theta (\bx,\by)
&=&\frac{1}{\ep^2}\int^\infty_z
\cF^{-1}_2\theta(\bx,\by+ik^{-1}
\ep^{-2\alpha}(s-z)\nabla)
\exp{(ik^{-1}\ep^{-2\alpha} (s-z)\Delta/2)}\\
&&\quad\lt[\IE^\ep_z\lt[V_s^\ep(\frac{\bx}{\ep^{2\alpha}}
-\frac{\ep^{2-2\alpha}\by}{2})\rt]-\IE^\ep_z\lt[V_s^\ep(\frac{\bx}{\ep^{2\alpha}}
+\frac{\ep^{2-2\alpha}\by}{2})\rt]\rt]ds.\nn
\eeq
****************}

The rest of the argument for tightness proceeds without changes.

\subsection{Identification of the limit}
With straightforward modification on the estimates of
the remainder terms $R_2^\ep$ and $R_3^\ep$ the same
argument for passing to the limit
$\ep\to 0$ as before carries over here.

In particular, Proposition~\ref{quad} can be
proved as follows.
\begin{prop}
\label{quad2}
\[
\lim_{\ep\to 0}\sup_{\|\psi\|_2=1}
A_2^{(1)}(\psi)
=0.
\]
\end{prop}
\begin{proof}
The kernel $\cQ_1(\theta\otimes\theta)$ has the
following expressions
\beqn
\lefteqn{\cQ_1(\theta\otimes\theta)(\bx,\bp,\by,\bq)}\\
&=&
\int^\infty_0
\int
\check{\Phi}(s,\bp')
e^{i\bp'\cdot(\bx-\by)/\ep^{2\alpha}}e^{-ik^{-1} s\bp\cdot\bp'\ep^{2-2\alpha}}
\ep^{2\alpha-2}\lt[\theta(\bx,\bp+\ep^{2-2\alpha}\bp'/2)-\theta(\bx,\bp-
\ep^{2-2\alpha}\bp'/2)\rt]\\
&&\times\ep^{2\alpha-2}\lt[\theta(\by,\bq+\ep^{2-2\alpha}\bp'/2)-\theta(\by,\bq-
\ep^{2-2\alpha}\bp'/2)\rt]\,\,d\bp'\,\,ds\\
&=&
\lim_{\ep\to0}
\pi\int
\Phi(k^{-1}\bp\cdot\bp'\ep^{2-2\alpha}, \bp')
e^{i\bp'\cdot(\bx-\by)/\ep^{2\alpha}}
\ep^{2\alpha-2}\lt[\theta(\bx,\bp+\ep^{2-2\alpha}\bp'/2)-\theta(\bx,\bp-
\ep^{2-2\alpha}\bp'/2)\rt]\\
&&\quad
\times
\ep^{2\alpha-2}
\lt[\theta(\by,\bq+\ep^{2-2\alpha}\bp'/2)-\theta(\by,\bq-\ep^{2-2\alpha}\bp'/2)\rt]\,\,d\bp'.
\eeqn
Note that
\[
\ep^{2\alpha-2}\lt[\theta(\bx,\bp+\ep^{2-2\alpha}\bp'/2)-\theta(\bx,\bp-
\ep^{2-2\alpha}\bp'/2)\rt]
\longrightarrow \bp'\cdot\nabla_\bp \theta(\bx,\bp)
\quad\hbox{in}\,\,C^\infty_c(\IR^{2d}).
\]
Thus the $L^2$-norm of $\cQ_1(\theta\otimes\theta)$
tends to zero for the same reason as
given in the proof of Proposition~\ref{quad}.
\end{proof}

To identify the limit we have the following
straightforward calculation: For any real-valued,
$L^2$-weakly convergent sequence $\psi^\ep\to\psi$
\beqn
\lefteqn{\lim_{\ep\to 0}
A_1^{(1)}(\psi^\ep)}\\
&=&
\lim_{\ep\to 0} \int^\infty_0ds\int dw
d\bq d\bx d\bp\,\, \psi^\ep(\bx,\bp)
\Phi(w,\bq) e^{isw}e^{-ik^{-1}s\bp\cdot\bq \ep^{2-2\alpha}}
\ep^{4\alpha-4}\\
&&\quad \lt[e^{-ik^{-1}s|\bq|^2\ep^{4-4\alpha}/2}
\lt[\theta(\bx,\bp+\ep^{2-2\alpha}\bq)-\theta(\bx,\bp)\rt]
-e^{ik^{-1}s|\bq|^2\ep^{4-4\alpha}/2}
\lt[\theta(\bx,\bp)-\theta(\bx,\bp-\ep^{2-2\alpha}\bq)\rt]
\rt]\\
&=&\lim_{\ep\to0}
\int^\infty_0ds \int
d\bq d\bx d\bp\,\, \psi^\ep(\bx,\bp)
\check{\Phi}(s,\bq)
e^{-ik^{-1}s\bp\cdot\bq \ep^{2-2\alpha}}\ep^{4\alpha-4}\\
&&\quad\lt[e^{-ik^{-1}s|\bq|^2\ep^{4-4\alpha}/2}
\lt[\theta(\bx,\bp+\ep^{2-2\alpha}\bq)-\theta(\bx,\bp)\rt]
-e^{ik^{-1}s|\bq|^2\ep^{4-4\alpha}/2}
\lt[\theta(\bx,\bp)-\theta(\bx,\bp-\ep^{2-2\alpha}\bq)\rt]\rt]\\
&=&
\pi\int d\bq d\bx d\bp\,\,\psi(\bx,\bp)
\Phi(0,\bq)(\bq\cdot\nabla_\bp)^2\theta(\bx,\bp)\\
&=&\bar{A}_1(\psi)
\eeqn
for $\alpha\in (0,1)$.
For $\alpha=1$ we have the same result as in
Theorem~1 Case (ii); for $\alpha>1$, the limit is
identically zero.

\section{Proof of Theorem~3}

The proof of the result for Case (i) and (iii)
is identical to that for Theorem~1, Case (i) and (iii),
respectively. So in the sequel we focus on the
second case $\alpha>\beta$.

Introducing a new parameter
\[
\tilde{\ep}=\ep^\beta
\]
we can rewrite the equation as
\beq
\label{beta3'}
\frac{\partial \wep_z}{\partial z}
+\frac{\bp}{\ks}\cdot\nabla \wep_z
+\frac{\ks}{\eptil}\cv \wep_z=0
\eeq
with
\beq
\label{L3.2'}
\lefteqn{\cv\wep_z(\bx,\xtil,\bp)}\\
&=&
i\eptil^{1-\al/\beta}\int e^{i\bq\cdot\xtil}
\lt[\wep_z(\bx,\bp+\bq/2)-\wep_z(\bx,\bp-\bq/2)\rt]
\widehat{V}_z^\ep(d\bq),\quad\xtil=\bx\eptil^{-2\al/\beta},\,\,
\beta<1
\nn
\eeq
with
\beq
\label{n21}
\widehat{V}_z^\ep(d\bq)=\widehat{V}(\frac{z}{\eptil^2}, d\bq),\quad
\vep(\bx)=V(\frac{z}{\eptil^2},\bx).
\eeq
Note again that
\beq
\invf\cv\theta=
-i\eptil^{1-\al/\beta}\delta_\ep V^\ep_z(\xtil,\by)
\invf\theta
\eeq
with
\beq
\label{n20}
\delta_\ep \vep (\xtil,\by)=
\vep(\xtil+\by/2)-\vep(\xtil-\by/2).
\eeq
We will work with (\ref{L3.2'}) and
(\ref{n21}) and construct the perturbed test
function in the power of $\eptil$.

First we note that
\beq
\IE\lt[\cv\theta\rt]^2(\bx,\bp)
&=&\eptil^{2-2\al/\beta}
\int [\theta(\bx,\bp+\bq/2)-\theta(\bx,\bp-\bq/2)]^2
\Phi(\xi, \bq)d\xi d\bq
\eeq
which has an $\ep$-uniformly bounded
support.

Instead of (\ref{mix}) we define
\beq
\label{mix3}
{\tilde{\cL}^\ep_z\theta(\bx,\tilde{\bx},\bp)}
&=&
\frac{i}{\eptil^{1+\al/\beta}}
\int_z^\infty
\int e^{i\bq\cdot\xtil}e^{i
k^{-1}(s-z)\bp\cdot\bq/\eptil^{2\al/\beta}}
[\theta(\bx,\bp+\bq/2)-
\theta(\bx,\bp-\bq/2)]\IE_z^\ep\hat{V}^\ep_s(d\bq)
\nn
\eeq
with
\[
\xtil=\bx\eptil^{-2\alpha/\beta}.
\]
After the partial inverse Fourier transform 
\beq
\invf\cvtil\theta(\bx,\xtil,\by)
=\frac{i}{\eptil^{1+\al/\beta}}
\int^\infty_z e^{ik^{-1}(s-z)\gry\cdot\grxtil
\eptil^{-2\al/\beta}}\lt[
\IE^\ep_z[\delta_\ep V^\ep_s] \invf
\theta\rt](\bx,\xtil,\by)ds\label{n52}
\eeq
 The
corrector equation holds again:
\beq
\label{imp3}
\lt[\eptil^{-2\al/\beta}\frac{\bp}{k}\cdot\nabla_{\xtil}+\cA^\ep\rt]
\tilde{\cL}^\ep_z\theta=\eptil^{-2}\cL^\ep_z\theta.
\eeq

Following the same argument as in the proof
of Theorem~1 we have the following estimates:
\begin{lemma}
\label{Lemma5}
\beq
\label{n22}
\IE\lt[\cltil_z\theta\rt]^2(\bx,\bp) \leq
\eptil^{2-2\al/\beta}\lt[\int^\infty_0\rho(s)ds \rt]^2
\int \lt[\theta(\bx,\bp+\bq/2)-\theta(\bx,\bp-\bq/2)\rt]^2
\Phi(\xi,\bq)d\xi d\bq
\eeq
which has an $\ep$-uniformly bounded
support.
\end{lemma}
\begin{cor}
\label{cor5}
\beq
\label{n23}
\lefteqn{\IE\lt[\bp\cdot\nabla_\bx\cltil_z\theta\rt]^2(\bx,\bp)}\\
& \leq& \eptil^{2-2\al/\beta}
\lt[\int^\infty_0\rho(s)ds \rt]^2
\int \lt[\bp\cdot\nabla_\bx\theta(\bx,\bp+\bq/2)-\bp\cdot\nabla_\bx\theta(\bx,\bp-\bq/2)\rt]^2
\Phi(\xi,\bq)d\xi d\bq\nn
\eeq
which has an $\ep$-uniformly bounded
support.
\end{cor}

\begin{lemma}
\label{Lemma6}
\beq
\IE\|\cv\cvtil\theta\|_2^2
&\leq
&\eptil^{4-4\al/\beta}8C\lt(\int^\infty_0\rho(s)ds\rt)^2
 \IE[ V_z]^2
  \int 
    [\theta(\bx,\bp+\bq/2)-
       \theta(\bx,\bp-\bq/2)]^2\Phi(\xi, \bq)d\xi d\bx d\bq d\bp
       \nn\\
\IE\|\cvtil\cvtil\theta\|_2^2
&\leq
&\eptil^{4-4\al/\beta}8C\lt(\int^\infty_0\rho(s)ds\rt)^4
 \IE[ V_z]^2
  \int 
    [\theta(\bx,\bp+\bq/2)-
       \theta(\bx,\bp-\bq/2)]^2\Phi(\xi, \bq)d\xi d\bx d\bq d\bp
       \nn
       \eeq
       for some constant $C$ independent of $\ep$.
\end{lemma}
\begin{cor}
\label{cor6}
\beqn
\lefteqn{\IE\|\bp\cdot\nabla_\bx\cltil_z\cltil_z\theta\|_2^2}\\
&\leq&\eptil^{4-4\al/\beta}
32C\lt(\int^\infty_0\rho(s)ds\rt)^4\lt\{
 \IE[\nabla_\by V_z]^2
 \int 
  [\nabla_\bx\theta(\bx,\bp+\bq/2)-
   \nabla_\bx\theta(\bx,\bp-\bq/2)]^2\Phi(\xi, \bq)d\xi d\bx
d\bq d\bp\rt.\\
&&\hspace{3cm}+\lt.
 \IE[V_z]^2
 \int 
  [\nabla_\bx\theta(\bx,\bp+\bq/2)-
   \nabla_\bx\theta(\bx,\bp-\bq/2)]^2|\bp|^2\Phi(\xi,
\bq)d\xi d\bx d\bq d\bp\rt\};
\eeqn
\beqn
\lefteqn{\IE\|\cv\cltil_z\cltil_z\theta\|_2^2}\\
&\leq&\eptil^{6-6\al/\beta}
32C\lt(\int^\infty_0\rho(s)ds\rt)^4
 \IE[ V_z]^4
 \int 
  [\theta(\bx,\bp+\bq/2)-
   \theta(\bx,\bp-\bq/2)]^2\Phi(\xi, \bq)d\xi d\bx d\bq d\bp
\eeqn
for some constant $C$ independent of $\ep$.
\end{cor}

As a consequence of the divergent factor
$\eptil^{1-1/\beta}$ in the above estimates
the previous proof of uniform integrability
of $\cA^\ep [f(z)+\eptil f^\ep_1]$ (e.g.
Proposition~\ref{prop:1}) breaks down.
For both the tightness and the identification we shall
use the test function
\[
f^\ep(z)=f(z)+f^\ep_1(z)+f_2^\ep(z)+
f_3^\ep(z)
\]
where 
\beq
f_1^\ep(z)&=&k\eptil f'(z)\lan \wepz,\cvtil\theta\ran\\
\label{2nd'}
f_2^\vas(z) &=&\frac{\eptil^2\ks^2}{2}f''(z)
\lt[\lan\psiep,\cvtil\theta\ran^2-A^{(2)}_2(\psiep)\rt]\\
f_3^\vas(z) &=&
\frac{\eptil^2\ks^2}{2}f'(z)\lt[\lan\psiep,\cvtil\cvtil\theta\ran-
A^{(2)}_1(\psiep)\rt].
\label{3rd'}
\eeq
with $A^{(2)}_1,A^{(2)}_2$ given as before.

Following the same procedure as in the proof of
Theorem~1
we obtain
\beq
\label{n29}
\lefteqn{\cA^\ep f^\ep(z)}\\
&=&
\frac{1}{\ks}f'(z)\lan\psiep,\pdgx\theta\ran+{\ks^2}f''(z)
A_2^{(1)}(\psiep)+{\ks^2}f'A_1^{(1)}(\psiep)
+R_2^\ep(z)+R_3^\ep(z)+A_3^\ep(z).\nn
\eeq
where
\beq
\label{n25}
R_2^\vas(z)& =&\eptil^2{\ks^2}\frac{f'''(z)}{2}\left[
\frac{1}{\ks}
\lan\psiep,\pdgx\theta\ran+\frac{\ks}{\eptil}\lan\psiep,\cv\theta\ran\rt]
\lt[\lan\psiep,\cvtil\theta\ran^2-A_2^{(2)}(\psiep)\rt]\\
\nn
&& +
\eptil^2{\ks^2}f''(z)\lan\psiep,\cvtil\theta\ran\lt[
\frac{1}{\ks}\lan\psiep,\pdgx(\cvtil\theta)\ran+\frac{\ks}{\eptil}
\lan\psiep,\cv\cvtil\theta\ran\rt]\\
&&-\eptil^2{\ks^2}f''(z)
\lt[\frac{1}{\ks}\lan\psiep,\pdgx(G_\theta^{(2)}\psiep)\ran+
\frac{\ks}{\eptil}\lan\psiep,\cv
G_\theta^{(2)}\psiep\ran\rt]
\nn
\eeq
\beq
\label{n26}
R^\ep_3(z)&=&\eptil^2{\ks^2}
f'(z)\left[\frac{1}{\ks}\lan\psiep,\pdgx(\cvtil\cvtil\theta)\ran+\frac{\ks}{\eptil}
\lan\psiep,\cv\cvtil\cvtil\theta\ran\rt]\\
&&\nn +\eptil^2\frac{\ks^2}{2}
f''(z)\left[
\frac{1}{\ks}\lan\psiep,\pdgx\theta\ran+\frac{\ks}{\eptil}\lan\psiep,\cv\theta\ran\rt]
\lt[\lan\psiep,\cvtil\cvtil\theta\ran-A_1^{(2)}(\psiep)\rt]\\
&&\nn-\eptil^2{\ks^2}f'(z)\lt[
\frac{1}{\ks}\lan\psiep,\pdgx(\cQ_2'\theta)\ran+\frac{\ks}{\eptil}
\lan\psiep,\cv\cQ_2'\theta\ran\rt]
\eeq
and $A_3^\ep, A^{(1)}_2, A^{(1)}_1, G_\theta^{(2)},
\cQ_2'$ all have the same expressions as in the proof
of Theorem~1.

With the assumption $\beta>3/4$,
Propositions~\ref{prop:2}, \ref{prop:3}, \ref{prop:4}
and \ref{prop:5} hold true. Let us remark
that the most severe terms due to
the divergent factor $\eptil^{1-\al/\beta}$ are
\beq
\label{n55}
\sup_{z<z_0}\eptil \IE\lt|\lan\psiep,\cv G_\theta^{(2)}
\psiep\ran\rt|&=&O(\eptil^{4-3\al/\beta}),\\
{\sup_{z<z_0}\eptil
\IE\lt|\lan\psiep,\cv\cvtil\cvtil\theta\ran\rt|}
&=&O(\eptil^{4-3\al/\beta})
\label{n56}
\eeq
(cf. Corollary~\ref{cor6}).

To satisfy (\ref{n59'}) we need to
show 
\begin{prop}\label{prop:8}
$$ \lim_{\ep\to
0}\sup_{z<z_0} |f_{j,z}^\vas|=0,
\quad j=2,3,  \quad\hbox{in
probability}.
$$
\end{prop}
\begin{proof}
 We
have the estimates
\beqn
\sup_{z<z_0}|f_{2,z}^\vas|&\leq&
\sup_{z<z_0}{\tilde{\vas}^2\ks^2}\|f''\|_\infty\lt[
\| W_0\|_2^2\|\cvtil\theta\|_2^2
+A^{(2)}_2(\psiep)\rt]\\
\sup_{z<z_0}
|f_{3,z}^\vas|&\leq&
\sup_{z<z_0}
{\tilde{\vas}^2\ks^2}\|f'\|_\infty
\lt[\| W_0\|_2\|\cvtil\cvtil\theta\|_2
+A_1^{(2)}(\psiep)\rt]
\eeqn
\commentout{****
and
\beq
\sup_{z<z_0}\|\cvtil\theta\|_2^2&\leq&\\
\|\cvtil\cvtil\theta\|_2&\leq&
\eeq
****}
which vanish in probability by using 
Assumption~3, Lemma~\ref{Lemma5}, \ref{Lemma6}
and  Chebyshev's inequality.
 
\end{proof}

\begin{prop}
\label{quad3}
\[
\lim_{\ep\to 0}\sup_{z<z_0}\sup_{\|\psi\|_2=1}
A_2^{(1)}(\psi)
=0.
\]
\end{prop}
\begin{proof}
The kernel $\cQ_1(\theta\otimes\theta)$ has the
following expressions in terms of $\ep$ (not $\eptil$):
\beqn
\lefteqn{\cQ_1(\theta\otimes\theta)(\bx,\bp,\by,\bq)}\\
&=&
\ep^{2\beta-2\al}\int^\infty_0
\int
\check{\Phi}(s,\bp')
e^{i\bp'\cdot(\xtil-\ytil)}e^{-ik^{-1}
s\bp\cdot\bp'\ep^{2\beta-2\al}}
\lt[\theta(\bx,\bp+\bp'/2)-\theta(\bx,\bp-
\bp'/2)\rt]\\
&&\times\lt[\theta(\by,\bq+\bp'/2)-\theta(\by,\bq-
\bp'/2)\rt]\,\,d\bp'\,\,ds\\
&=& \ep^{2\beta-2\al}
\pi\int
\Phi(k^{-1}\bp\cdot\bp'\ep^{2\beta-2\al}, \bp')
e^{i\bp'\cdot(\xtil-\ytil)}
\lt[\theta(\bx,\bp+\bp'/2)-\theta(\bx,\bp-
\bp'/2)\rt]\\
&&\quad
\times
\lt[\theta(\by,\bq+\bp'/2)-\theta(\by,\bq-\bp'/2)\rt]\,\,d\bp'.
\eeqn
Taking the $L^2$-norm and passing to the limit we have
\beqn
&&\lim_{\ep\to 0}\int
d\bx d\bp d\by d\bq \left|
{\cQ_1(\theta\otimes\theta)(\bx,\bp,\by,\bq)}\right|^2\\
&=&\lim_{\ep\to}
\pi^2\int
d\bx d\bp d\by d\bq 
\left|\int \delta(\frac{\bp\cdot\bp'}{k})\lt[\int
\Phi(w,\bp')dw\rt]
e^{i\bp'\cdot(\bx-\by)/\ep^2}\right.\\
&&\quad\times
\left.\lt[\theta(\bx,\bp+\bp'/2)-\theta(\bx,\bp-
\bp'/2)\rt]\lt[\theta(\by,\bq+\bp'/2)-\theta(\by,\bq-
\bp'/2)\rt]\,\,d\bp'\right|^2\\
&=&\lim_{\ep\to 0}
\pi^2\int
d\bx d\bp d\by d\bq 
\left|\int\,\,\frac{k}{|\bp|}
\lt[\int
\Phi(w,\bp_\perp)dw\rt]
 e^{i\bp_\perp\cdot(\bx-\by)/\ep^2}\right.\\
 &&\quad\times
 \left.
\lt[\theta(\bx,\bp+\bp_\perp/2)-\theta(\bx,\bp-
\bp_\perp/2)\rt]\lt[\theta(\by,\bq+\bp_\perp/2)-\theta(\by,\bq-
\bp_\perp/2)\rt]\,\,d\bp_\perp\right|^2
\eeqn
where $\bp_\perp \cdot\bp=0, \bp_\perp\in \IR^{d-1}$.
In passing to the limit
the only problem is at the point $\bp=0$. But the 
integrand in the above integral 
is bounded by
\beq
\label{dom}
C|\bp|^{-2},\quad\hbox{some constant } C<\infty
\eeq
which is integrable in a neighborhood of
zero if $d\geq 3$. 
Hence the $L^2$-norm of $\cQ_1(\theta\otimes\theta)$
tends to zero by
the dominated convergence theorem.
\end{proof}

We have the following
straightforward calculation:
For any real-valued, $L^2$-weakly convergent sequence
$\psi^\ep\to
\psi$,
\beqn
\lefteqn{\lim_{\ep\to 0}
A_1^{(1)}(\psi^\ep)}\\
&=&
\lim_{\ep\to 0} \int^\infty_0ds  \int
d\bq d\bx d\bp\,\, \psi^\ep(\bx,\bp)
\check{\Phi}(s,\bq)\ep^{2\beta-2\al}\\
&&\lt[e^{-ik^{-1}s(\bp+\bq/2)\cdot\bq 
\ep^{2\beta-2\al}}
\lt[\theta(\bx,\bp+\bq)-\theta(\bx,\bp)\rt]
-e^{-ik^{-1}s(\bp-\bq/2)\cdot\bq\ep^{2\beta-2\al}}
\lt[\theta(\bx,\bp)-\theta(\bx,\bp-\bq)\rt]
\rt]\\
&=&\lim_{\ep\to0}
\pi\int
d\bq d\bx d\bp\,\, \psi^\ep(\bx,\bp)
\lt[{\Phi}(k^{-1}(\bp+\bq/2)\cdot\bq\ep^{2\beta-2\al},\bq)\ep^{2\beta-2\al}
\lt[\theta(\bx,\bp+\bq)-\theta(\bx,\bp)\rt]\right.\\
&&\quad\left.-\Phi(k^{-1}(\bp-\bq/2)\cdot\bq\ep^{2\beta-
2\al},\bq)\ep^{2\beta-2\al}
\lt[\theta(\bx,\bp)-\theta(\bx,\bp-\bq)\rt]\rt]\\
&=&
\lim_{\ep\to0}
\pi\int
d\bq d\bx d\bp\,\, \psi^\ep(\bx,\bp)
\lt[{\Phi}(\frac{|\bq|^2-|\bp|^2}{2k}\ep^{2\beta-2\al},\bq-\bp)\ep^{2\beta
-2\al}
\lt[\theta(\bx,\bp+\bq)-\theta(\bx,\bp)\rt]\right.\\
&&\quad\left.-\Phi(\frac{|\bq|^2-|\bp|^2}{2k}\ep^{2\beta
-2\al},\bq-\bp)\ep^{2\beta-2\al}
\lt[\theta(\bx,\bp)-\theta(\bx,\bp-\bq)\rt]\rt]\\
&=&2\pi\int
d\bq d\bx d\bp\,\, \psi(\bx,\bp)
\delta(\frac{|\bq|^2-|\bp|^2}{2k})\lt[\int\Phi(w,\bq-\bp)dw\rt]
\lt[\theta(\bx,\bp+\bq)-\theta(\bx,\bp)\rt]\\
&\equiv&\bar{A}_1(\psi)
\eeqn
following from the strong convergence of
\beqn
&&\int 
\lt[{\Phi}(\frac{|\bq|^2-|\bp|^2}{2k}\ep^{2\beta-2\al},\bq-\bp)\ep^{2\beta-2\al}
\lt[\theta(\bx,\bp+\bq)-\theta(\bx,\bp)\rt]\right.\\
&&\quad\left.-\Phi(\frac{|\bq|^2-|\bp|^2}{2k}\ep^{2\beta-
2\al},\bq-\bp)\ep^{2\beta-2\al}
\lt[\theta(\bx,\bp)-\theta(\bx,\bp-\bq)\rt]\rt]d\bq,\quad
\forall \Phi \in C^\infty_c(\IR^{d+1}),\theta\in C^\infty_c(\IR^{2d})
\eeqn
in $L^2(\IR^{2d})$.

\begin{prop}
$\cA^\ep f^\ep(z)$ is uniformly integrable.
\end{prop}
This, of course, follows from the fact
that each term in (\ref{n29}) has a uniformly
bounded second moment. Therefore we have
completed the tightness argument.
Moreover, we have also identified
 the limiting equation.

\section{Proof of Theorem~4}
Introducing a new parameter
\[
\tilde{\ep}=\ep^\beta,
\]
the fast variable
\[
\xtil=\bx \epal
\]
and the rescaled process
\beq
\label{n31}
\widehat{V}_z^\ep(d\bq)=\widehat{V}(\frac{z}{\eptil^2}, d\bq),\quad
\vep(\bx)=V(\frac{z}{\eptil^2},\bx)
\eeq
we rewrite the equation as
\beq
\label{beta4'}
\frac{\partial \wep_z}{\partial z}
+\frac{\bp}{\ks}\cdot\nabla \wep_z
+\frac{\ks}{\eptil}\cv \wep_z=0
\eeq
with, in the case (i),
\beq
\label{L4.1}
\lefteqn{\cv\wep_z(\bx,\xtil,\bp)}\\
&=&
i\int e^{i\bq\cdot\xtil}
\ep^{2\alpha-2}\lt[\wep_z(\bx,\bp+\ep^{2-2\alpha}
\bq/2)-\wep_z(\bx,\bp-\ep^{2-2\alpha}\bq/2)\rt]
\widehat{V}_z^\ep(d\bq),
\nn
\eeq
and in the case (ii)
\beq
\label{L4.2}
\lefteqn{\cv\wep_z(\bx,\xtil,\bp)}\\
&=&
i\eptil^{1-\alpha/\beta}\int e^{i\bq\cdot\xtil}
\ep^{2\alpha-2}\lt[\wep_z(\bx,\bp+\ep^{2-2\alpha}
\bq/2)-\wep_z(\bx,\bp-\ep^{2-2\alpha}\bq/2)\rt]
\widehat{V}_z^\ep(d\bq).
\nn
\eeq 

Taking the partial inverse Fourier transform we get
in the case (i) 
\beq
\invf\cv\theta(\bx,\xtil,\by)=
-i\delta_\ep V^\ep_z(\xtil,\by)
\invf\theta(\bx,\by)
\label{n35}
\eeq
and in the case (ii)
\beq
\invf\cv\theta(\bx,\xtil, \by)=
-i\eptil^{1-\alpha/\beta}\delta_\ep V^\ep_z(\xtil,\by)
\invf\theta(\bx,\by)
\label{n37}
\eeq
with
\beq
\label{n36}
\delta_\ep \vep (\xtil,\by)=
\ep^{2\alpha-2}\lt[\vep(\xtil+\ep^{2-2\alpha}\by/2)-\vep(\xtil-
\ep^{2-2\alpha}\by/2\rt].
\eeq

The proof for the case (i) is entirely analogous to
that for Theorem~2 and we will focus on the case (ii)
in the sequel. And we will work with 
(\ref{L4.2}) and (\ref{n37}) and construct the
perturbed test function in the power of $\eptil$.

First we note that
\beq
\limsup_{\ep\to 0} \eptil^{-2+2\alpha/\beta}\IE\lt[\cv\theta\rt]^2(\bx,\bp)
&=&
\int [\bq\cdot\nabla_\bp\theta(\bx,\bp)]^2
\Phi(\xi, \bq)d\xi d\bq
\eeq
which has an $\ep$-uniformly bounded
support.

As in the proof of Theorem~3, we carry the analysis
in the power of $\tilde{\ep}=\ep^{\beta}$.
We consider the rescaled process (\ref{n21})
and its sigma algebras.

We introduce the corrector
\beq
\nn\label{mix4}
\lefteqn{\tilde{\cL}^\ep_z\theta(\bx,\tilde{\bx},\bp)}\\
&=&
\frac{i}{\eptil^{1+\alpha/\beta}}
\int_z^\infty
\int e^{i\bq\cdot\xtil}e^{i k^{-1}(s-z)\bp\cdot\bq/\eptil^{2\alpha/\beta}}
\ep^{2\alpha-2}[\theta(\bx,\bp+\ep^{2-2\alpha}\bq/2)-
\theta(\bx,\bp-\ep^{2-2\alpha}\bq/2)]\IE_z^\ep\hat{V}^\ep_s(d\bq)
\nn
\eeq
which after the partial Fourier inversion becomes
\beq
\label{mix4.2}
\label{n53}
\invf\tilde{\cL}^\ep_z\theta(\bx,\tilde{\bx},\by)
&=&-
\frac{i}{\eptil^{1+\alpha/\beta}}
\int_z^\infty
e^{i k^{-1}(s-z)\gry\cdot\grxtil/\eptil^{2\alpha/\beta}}
\IE_z^\ep[\delta_\ep V^\ep_s]\invf \theta (\bx,\xtil,\by)ds.
\eeq
The corrector solves the corrector equation
eq. (\ref{imp3}). 

Following the same argument as in the proof
of Theorem~1 we have the following estimates:
\begin{lemma}
\label{Lemma7}
\beq
\label{n42}
\limsup_{\eptil\to 0} \eptil^{-2+2\alpha/\beta}
\IE\lt[\cltil_z\theta\rt]^2(\bx,\bp) \leq
\lt[\int^\infty_0\rho(s)ds \rt]^2
\int \lt[\bq\cdot\grp\theta(\bx,\bp)\rt]^2
\Phi(\xi,\bq)d\xi d\bq
\eeq
which has a compact
support.
\end{lemma}
\begin{cor}
\label{cor7}
\beq
\label{n43}
{\limsup_{\eptil\to 0} \eptil^{-2+2\alpha/\beta}
\IE\lt[\bp\cdot\nabla_\bx\cltil_z\theta\rt]^2(\bx,\bp)}
& \leq& 
\lt[\int^\infty_0\rho(s)ds \rt]^2
\int \lt[\bp\cdot\nabla_\bx\bq\cdot\grp
\theta(\bx,\bp)\rt]^2
\Phi(\xi,\bq)d\xi d\bq\nn
\eeq
which has a compact
support.
\end{cor}

\begin{lemma}
\label{Lemma8}
\beq
\limsup_{\eptil\to 0} \eptil^{-4+4\alpha/\beta}\IE\|\cv\cvtil\theta\|_2^2
&\leq
&8C\lt(\int^\infty_0\rho(s)ds\rt)^2
 \IE[ V_z]^2
   \int
       [\bq\cdot\grp\theta(\bx,\bp)
              ]^2\Phi(\xi, \bq)d\xi d\bx d\bq d\bp
	             \nn\\
		     \limsup_{\eptil\to 0}
		     \eptil^{-4+4\alpha/\beta}\IE\|\cvtil\cvtil\theta\|_2^2
		     &\leq
		     &8C\lt(\int^\infty_0\rho(s)ds\rt)^4
		      \IE[ V_z]^2
		        \int
    [\bq\cdot\grp\theta(\bx,\bp)
          ]^2\Phi(\xi, \bq)d\xi d\bx d\bq d\bp
		          \nn
        \eeq
       for some constant $C$ independent of $\ep$.
\end{lemma}
\begin{cor}
\label{cor8}
\beqn
\lefteqn{\limsup_{\eptil\to 0} \eptil^{-4+4\alpha/\beta}
\IE\|\bp\cdot\nabla_\bx\cltil_z\cltil_z\theta\|_2^2}\\
&\leq&
32C\lt(\int^\infty_0\rho(s)ds\rt)^4\lt\{
 \IE[\nabla_\by V_z]^2
  \int
    [\nabla_\bx \bq\cdot\grp\theta(\bx,\bp)
       ]^2\Phi(\xi, \bq)d\xi d\bx
       d\bq d\bp\rt.\\
       &&\hspace{3cm}+\lt.
        \IE[V_z]^2
	 \int
	   [\nabla_\bx\bq\cdot\nabla_\bp\theta(\bx,\bp)
	      ]^2|\bp|^2\Phi(\xi,
	      \bq)d\xi d\bx d\bq d\bp\rt\};
	      \eeqn
	      \beqn
	      \lefteqn{\limsup_{\eptil\to 0}\eptil^{-6+6\alpha/\beta}
	      \IE\|\cv\cltil_z\cltil_z\theta\|_2^2}\\
	      &\leq&
	      32C\lt(\int^\infty_0\rho(s)ds\rt)^4
	       \IE[ V_z]^4
	        \int
		  [\bq\cdot\grp\theta(\bx,\bp)
		     ]^2\Phi(\xi, \bq)d\xi d\bx d\bq d\bp
		     \eeqn
		     for some constant $C$ independent of $\ep$.
		     \end{cor}

The rest of the argument follows the general outline
of that of Theorem~3 Case (ii).

Let us know verify that the quadratic
variation vanishes in the limit.
\begin{prop}
\label{quad4}
\[
\lim_{\ep\to 0}\sup_{z<z_0}\sup_{\|\psi\|_2=1}
A_2^{(1)}(\psi)
=0.
\]
\end{prop}
\begin{proof}
\commentout{***********************************
{\bf Case (i): $\beta>\alpha$.}
The kernel $\cQ_1(\theta\otimes\theta)$  can be calculated as
follows.
\beqn
\lefteqn{\cQ_1(\theta\otimes\theta)(\bx,\bp,\by,\bq)}\\
&=&
\int^\infty_0
\int
\check{\Phi}(s,\bp')
e^{i\bp'\cdot(\bx-\by)/\ep^{2\alpha}}e^{-ik^{-1} s\bp\cdot\bp'\ep^{2\beta-2\alpha}}
\ep^{2\alpha-2}\lt[\theta(\bx,\bp+\ep^{2-2\alpha}\bp'/2)-\theta(\bx,\bp-
\ep^{2-2\alpha}\bp'/2)\rt]\\
&&\times\ep^{2\alpha-2}\lt[\theta(\by,\bq+\ep^{2-2\alpha}\bp'/2)-\theta(\by,\bq-
\ep^{2-2\alpha}\bp'/2)\rt]\,\,d\bp'\,\,ds\\
&=&
\pi\int
\Phi(k^{-1}\bp\cdot\bp'\ep^{2\beta-2\alpha}, \bp')
e^{i\bp'\cdot(\bx-\by)/\ep^{2\alpha}}
\ep^{2\alpha-2}\lt[\theta(\bx,\bp+\ep^{2-2\alpha}\bp'/2)-\theta(\bx,\bp-
\ep^{2-2\alpha}\bp'/2)\rt]\\
&&\quad
\times
\ep^{2\alpha-2}
\lt[\theta(\by,\bq+\ep^{2-2\alpha}\bp'/2)-\theta(\by,\bq-\ep^{2-2\alpha}\bp'/2)\rt]\,\,d\bp'.
\eeqn
Thus the $L^2$-norm of $\cQ_1(\theta\otimes\theta)$
tends to zero as 
in the proof of Proposition~\ref{quad2}.

{\bf Case (ii): $\beta<\alpha$.}
***************************************}
The kernel $\cQ_1(\theta\otimes\theta)$  can be calculated as
follows.
\beqn
\lefteqn{\cQ_1(\theta\otimes\theta)(\bx,\bp,\by,\bq)}\\
&=&
\ep^{2\beta-2\alpha}
\int^\infty_0
\int
\check{\Phi}(s,\bp')
e^{i\bp'\cdot(\bx-\by)/\ep^{2\alpha}}e^{-ik^{-1} s\bp\cdot\bp'\ep^{2\beta-2\alpha}}
\ep^{2\alpha-2}\lt[\theta(\bx,\bp+\ep^{2-2\alpha}\bp'/2)-\theta(\bx,\bp-
\ep^{2-2\alpha}\bp'/2)\rt]\\
&&\times\ep^{2\alpha-2}\lt[\theta(\by,\bq+\ep^{2-2\alpha}\bp'/2)-\theta(\by,\bq-\ep^{2-2\alpha}\bp'/2)\rt]\,\,d\bp'\,\,ds\\
&=&
\pi\int
\ep^{2\beta-2\alpha}\Phi(k^{-1}\bp\cdot\bp'\ep^{2\beta-2\alpha}, \bp')
e^{i\bp'\cdot(\bx-\by)/\ep^{2\alpha}}
\ep^{2\alpha-2}\lt[\theta(\bx,\bp+\ep^{2-2\alpha}\bp'/2)-\theta(\bx,\bp-
\ep^{2-2\alpha}\bp'/2)\rt]\\
&&\quad
\times
\ep^{2\alpha-2}
\lt[\theta(\by,\bq+\ep^{2-2\alpha}\bp'/2)-\theta(\by,\bq-\ep^{2-2\alpha}\bp'/2)\rt]\,\,d\bp'
\eeqn
whose $L^2$-norm has the following limit
\[
\lim_{\ep \to 0}
\pi^2\int\lt|\int_{\bp_\perp\cdot\bp=0}
k|\bp|^{-1} \lt[\int \Phi(w,\bp_\perp)dw\rt]
e^{i\bp_\perp\cdot(\bx-\by)\ep^{-2\alpha}}|\bq\cdot\nabla_\bp\theta(\bx,\bp)
|^2 d\bp_\perp \rt|^2 d\bx d\bp =0
\]
if $d\geq 3$ by the dominated convergence theorem because the
integrand is bounded by the integrable function
(\ref{dom}) in a neighborhood of zero.
\end{proof}

To identify the limit, we have the following
straightforward calculation: For any
real-valued, $L^2$-weakly convergent sequence
$\psi^\ep\to
\psi$,
\commentout{**************************************
{\bf Case (i): $\beta>\alpha$.}
\beqn
\lefteqn{\lim_{\ep\to 0}
A_1^{(1)}(\psi)}\\
&=&
\lim_{\ep\to 0} \int^\infty_0ds  \int
d\bq d\bx d\bp\,\, \psi(\bx,\bp)
\check{\Phi}(s,\bq)e^{-ik^{-1}s\bp\cdot\bq \ep^{2\beta-2\alpha}}
\ep^{4\alpha-4}\\
&&\quad \lt[e^{-ik^{-1}|\bq|^2\ep^{2+2\beta-4\alpha}/2}
\lt[\theta(\bx,\bp+\ep^{2-2\alpha}\bq)-\theta(\bx,\bp)\rt]
-e^{ik^{-1}|\bq|^2\ep^{2+2\beta-4\alpha}/2}
\lt[\theta(\bx,\bp)-\theta(\bx,\bp-\ep^{2-2\alpha}\bq)\rt]
\rt]\\
&=&\lim_{\ep\to0}
\int^\infty_0ds \int\int\int
d\bq d\bx d\bp\,\, \psi(\bx,\bp)
\check{\Phi}(s,\bq)
e^{-ik^{-1}s\bp\cdot\bq \ep^{2\beta-2\alpha}}\ep^{4\alpha-4}\\
&&\quad\lt[e^{-ik^{-1}s|\bq|^2\ep^{2+2\beta-4\alpha}/2}
\lt[\theta(\bx,\bp+\ep^{2-2\alpha}\bq)-\theta(\bx,\bp)\rt]
-e^{ik^{-1}|\bq|^2\ep^{4-4\alpha}/2}
\lt[\theta(\bx,\bp)-\theta(\bx,\bp-\ep^{2-2\alpha}\bq)\rt]\rt]\\
&=&\lim_{\ep\to0}
\pi\int\int\int
d\bq d\bx d\bp\,\, \psi(\bx,\bp)
\ep^{4\alpha-4}
\lt[{\Phi}(k^{-1}(\bp+\ep^{2-2\alpha}\bq/2)\cdot\bq \ep^{2\beta-2\alpha},\bq)
\lt[\theta(\bx,\bp+\ep^{2-2\alpha}\bq)-\theta(\bx,\bp)\rt]\right.\\
&&\quad-\left.
{\Phi}(k^{-1}(\bp-\ep^{2-2\alpha}\bq/2)\cdot\bq \ep^{2\beta-2\alpha},\bq)
\lt[\theta(\bx,\bp)-\theta(\bx,\bp-\ep^{2-2\alpha}\bq)\rt]\rt]\\
&=&
\pi\int\int\int d\bq d\bx d\bp\,\,\psi(\bx,\bp)
\Phi(0,\bq)(\bq\cdot\nabla_\bp)^2\theta(\bx,\bp),
\quad\forall \psi\in L^2(\IR^{2d}).
\eeqn

{\bf Case (ii): 
 $\beta<\alpha$.}
 ****************************}
 \beqn
 \lefteqn{\lim_{\ep\to 0}
 A_1^{(1)}(\psi^\ep)}\\
 &=&
 \lim_{\ep\to 0} \int^\infty_0ds \int
 d\bq d\bx d\bp\,\, \psi^\ep(\bx,\bp)
 \check{\Phi}(s,\bq)e^{-ik^{-1}s\bp\cdot\bq \ep^{2\beta-2\alpha}}
 \ep^{2\beta-2\alpha}\ep^{4\alpha-4}\\
 &&\quad
\lt[e^{-ik^{-1}s|\bq|^2\ep^{2+2\beta-4\alpha}/2}
 \lt[\theta(\bx,\bp+\ep^{2-2\alpha}\bq)-\theta(\bx,\bp)\rt]\right.\\
 &&\quad\left.
-e^{ik^{-1}s|\bq|^2\ep^{2+2\beta-4\alpha}/2}
 \lt[\theta(\bx,\bp)-\theta(\bx,\bp-\ep^{2-2\alpha}\bq)\rt]
 \rt]\\
 &=&\lim_{\ep\to0}
 \int^\infty_0ds \int
 d\bq d\bx d\bp\,\, \psi^\ep(\bx,\bp)
 \check{\Phi}(s,\bq)
 e^{-ik^{-1}s\bp\cdot\bq \ep^{2\beta-2\alpha}} \ep^{2\beta-2\alpha}
 \ep^{4\alpha-4}\\
 &&\quad\lt[e^{-ik^{-1}s|\bq|^2\ep^{2+2\beta-4\alpha}/2}
 \lt[\theta(\bx,\bp+\ep^{2-2\alpha}\bq)-\theta(\bx,\bp)\rt]\right.\\
 &&\quad \left.
-e^{ik^{-1}s|\bq|^2\ep^{2+2\beta-4\alpha}/2}
 \lt[\theta(\bx,\bp)-\theta(\bx,\bp-\ep^{2-2\alpha}\bq)\rt]\rt]\\
 &=&
 \lim_{\ep\to0}
 \pi\int
 \ep^{2\alpha-2}\ep^{2\beta-2\alpha}
 \lt[{\Phi}(k^{-1}(\bp+\ep^{2-2\alpha}\bq/2)\cdot\bq 
 \ep^{2\beta-2\alpha},\bq)
 \ep^{2-2\alpha}
 \lt[\theta(\bx,\bp+\ep^{2-2\alpha}\bq)-\theta(\bx,\bp)\rt]
 \right.\\
 &&\quad\left. -
 {\Phi}(k^{-1}(\bp-\ep^{2-2\alpha}\bq/2)\cdot\bq
  \ep^{2\beta-2\alpha},\bq)
  \ep^{2-2\alpha}
  \lt[\theta(\bx,\bp)-\theta(\bx,\bp-\ep^{2-2\alpha}\bq)\rt]\rt]
   \psi^\ep(\bx,\bp) d\bq d\bx d\bp\\
  &=&\pi\int
      \bq\cdot\nabla_\bp\lt[\delta(k^{-1}\bp\cdot\bq)
        \lt[\int \Phi(w,\bq)dw\rt]\bq\cdot\nabla_\bp\rt]\theta(\bx,\bp)
	 \psi(\bx,\bp) d\bq d\bx d\bp\\
 &=&\pi\int
  k|\bp|^{-1}
  \lt[\int \Phi(w,\bp_\perp)dw\rt] (\bp_\perp\cdot\nabla_\bp)^2\theta(\bx,\bp)
   \psi(\bx,\bp) d\bp_\perp d\bx d\bp\\
   &\equiv& \bar{A}_1(\psi)
 \eeqn
 where $\bp_\perp\in \IR^{d-1}, \bp_\perp\cdot\bp=0$.

{\bf Case (iii): $\alpha=\beta$}.
 \beqn
  \lefteqn{\lim_{\ep\to 0}
   A_1^{(1)}(\psi^\ep)}\\
   &=&\lim_{\ep\to0}
   \pi\int
   d\bq d\bx d\bp\,\, \psi^\ep(\bx,\bp)
   \ep^{4\alpha-4}
   \lt[{\Phi}(k^{-1}(\bp+\ep^{2-2\alpha}\bq/2)\cdot\bq,\bq)
   \lt[\theta(\bx,\bp+\ep^{2-2\alpha}\bq)-\theta(\bx,\bp)\rt]\right.\\
   &&\quad-\left.
   {\Phi}(k^{-1}(\bp-\ep^{2-2\alpha}\bq/2)\cdot\bq,\bq)
   \lt[\theta(\bx,\bp)-\theta(\bx,\bp-\ep^{2-2\alpha}\bq)\rt]\rt]\\
   &=&
   \pi\int d\bq d\bx d\bp\,\,\psi(\bx,\bp)
   \bq\cdot\nabla_\bp\lt[\Phi(k^{-1}\bp\cdot\bq,\bq)\bq\cdot\nabla_\bp\rt]
   \theta(\bx,\bp)\\
   &\equiv&\bar{A}_1(\psi).
   \eeqn

\commentout{***************
\begin{appendix}

\commentout{********************************
\section{Proof of Lemma~\ref{Lemma1}}
Without loss of generality we set 
$\ep=1$. Note that in this case
\beq
\label{mix1}
\tilde{\cL}_z\theta(\bx,\bp)
&=&
i
\int e^{i\bq\cdot\bx}
[\theta(\bx,\bp+\bq/2)-
\theta(\bx,\bp-\bq/2)]
\int_z^\infty e^{i
k^{-1}(s-z)\bp\cdot\bq}\IE_z\hat{V}_s(d\bq) ds\nn
\eeq
and
\bean
\cF^{-1}_{2}\tilde{\cL}_z\theta (\bx,\by)
&=&\int \lt[ \tilde{V}_{z}
({\bx}-
\frac{\by}{2},\bp)-
\tilde{V}_{z}({\bx}+
\frac{\by}{2},\bp)\rt]e^{i\bp\cdot\by}
\theta(\bx,\bp)\,d\bp
\eean
with
\beq
\label{new2}
\tilde{V}_z(\bx,\bp)&=&
\int^\infty_z
e^{-k^{-1}(s-z)\bp\cdot\nabla}
e^{ik^{-1}(s-z)\Delta/2}
\IE_z\lt[ V_s(\bx)\rt]\,\,ds\\
&=&\int e^{i\bk\cdot\bx}\int^\infty_z
e^{ik^{-1}(s-z)|\bk|^2/2}
e^{-ik^{-1}(s-z)\bp\cdot\bk}
\IE_z\lt[
\hat{V}_s(d\bk)\rt]\,\,ds.
\nn
\eeq

By the identity 
\[
\lan \tilde{\cL}_z\theta,
\tilde{\cL}_z\theta\ran=
\lan\cF^{-1}_2 \tilde{\cL}_z\theta,\cF^{-1}_2
\tilde{\cL}_z\theta\ran
\]
we see that to prove the existence of the second moment it suffices to show
\beq
\int
\IE\lt[\tilde{V}_{z}({\bx}\pm
\frac{\by}{2},\bp)\overline{\tilde{V}_{z}({\bx}\pm
\frac{\by}{2},\bp')}\rt]e^{i\bp\cdot\by}e^{-i\bp'\cdot\by}
\theta(\bx,\bp) \theta(\bx,\bp')\,d\bp d\bp'
d\bx d\by<\infty.
\label{2.86}
\eeq
The left side of (\ref{2.86}) can be bounded
by 
\beqn
&&\int\int^\infty_z\int^\infty_z
e^{ik^{-1}(s-z)|\bk|^2/2}
e^{-ik^{-1}(t-z)|\bk'|^2/2}
e^{-ik^{-1}(s-z)\bp\cdot\bk}
e^{ik^{-1}(t-z)\bp'\cdot\bk'}e^{i(\bp-\bp')\cdot
\by}\\
&&\quad\theta(\bx,\bp)
\theta(\bx,\bp')e^{i\bk\cdot(\bx\pm
\by/2)} e^{i\bk'\cdot(\bx\pm \by/2)}
\IE\lt[\IE_z\lt[
\hat{V}_s(d\bk)\rt]
\IE_z\lt[\hat{V}_t(d\bk')\rt]\rt]\,\,ds  dt d\bx
d\by d\bp d\bp'\\
&=&\int\int^\infty_z\int^\infty_z
\cF^{-1}_2\theta(\bx, \by-k^{-1}(s-z)\bk)
\cF^{-1}_2\theta(\bx,\by-k^{-1}(t-z)\bk')\\
&&e^{i\bk\cdot(\bx\pm
\by/2)} e^{i\bk'\cdot(\bx\pm \by/2)}
e^{ik^{-1}(s-z)|\bk|^2/2}
e^{-ik^{-1}(t-z)|\bk'|^2/2}\IE\lt[\IE_z\lt[
\hat{V}_s(d\bk)\rt]
\IE_z\lt[\hat{V}_t(d\bk')\rt]\rt]\,\,ds  dt d\bx
d\by\\
&=&\int\IE\lt\{\int^\infty_z
\cF^{-1}_2\theta(\bx, \by-k^{-1}(s-z)\bk)
e^{i\bk\cdot(\bx\pm
\by/2)}
e^{ik^{-1}(s-z)|\bk|^2/2}
\IE_z\lt[
\hat{V}_s(d\bk)\rt]
\,\,ds \rt\}^2 d\bx
d\by\\
&=&\int\int^\infty_z \int^\infty_z\IE\lt\{\int
\cF^{-1}_2\theta(\bx, \by-k^{-1}(s-z)\bk)
e^{i\bk\cdot(\bx\pm
\by/2)}
e^{ik^{-1}(s-z)|\bk|^2/2}
\IE_z\lt[
\hat{V}_s(d\bk)\rt]
\right.\\
&&\lt. \int
\cF^{-1}_2\theta(\bx, \by-k^{-1}(t-z)\bk')
e^{i\bk'\cdot(\bx\pm
\by/2)}
e^{ik^{-1}(t-z)|\bk'|^2/2}
\hat{V}_t(d\bk')
\rt\}\,\, ds dt  d\bx
d\by
\commentout{
&=&\int\int^\infty_z\int^\infty_z
\cF^{-1}_2\theta(\bx, \by-k^{-1}(s-z)\bk)
\cF^{-1}_2\theta(\bx,\by-k^{-1}(t-z)\bk)\\
&&
e^{ik^{-1}(s-t)|\bk|^2/2}
\IE\lt[\IE_z\lt[
\hat{V}_s(d\bk)\rt]
\IE_z\lt[\hat{V}_t(d\bk')\rt]\rt]\,\,ds  dt d\bx
d\by
}
\eeqn

\commentout{****************************************************
 Consider the strong mixing coefficient
\beqn
\alpha(t)&=&\sup_{A\in \cF^+_{z+t}}\sup_{B\in \cF_z}
|P(AB)-P(A)P(B)|\\
&=&\frac{1}{2}\sup_{A\in \cF_{z+t}}\IE\lt[ |P(A|\cF_z)-P(A)|\rt]
\eeqn
and 
the general $L^p$-mixing coefficients
\beqn
\phi_p(t)
&=&\sup_{A\in \cF^+_{z+t}}\IE^{1/p}\lt[
|P(A|\cF_z)-P(A)|^p\rt],\quad p\in [1,\infty)\\
&=& \sup_{h\in L^p(P,\cF^+_{z+t})}\sup_{g\in L^q(P,
\cF_z)\atop
\IE g^q=1, \IE g=0} \IE[ hg],\quad p^{-1}+q^{-1}=1,\quad p\in
[1,\infty)
\eeqn
We note that $\alpha(t)=\phi_1(t)$ and for $p=\infty$
\beqn
\phi_\infty(t)&=&\sup_{A\in \cF^+_{t+z}}\sup_{B\in
\cF_z\atop P(B)>0}|P(A|B)-P(A)|, \quad \forall
t\geq 0\\ &=&\sup_{A\in
\cF^+_{t+z}}\hbox{ess-sup}_\omega|P(A|\cF_z)-P(A)|\\
&\equiv& f(t)
\eeqn
 is called
the uniform mixing coefficient \cite{EK}. 
***********************************************************}

Now take 
\beqn
h=h_s(\bx,\by)&=&\int
\cF^{-1}_2\theta(\bx, \by-k^{-1}(s-z)\bk)
e^{i\bk\cdot(\bx\pm
\by/2)}
e^{ik^{-1}(s-z)|\bk|^2/2}
\IE_z\lt[
\hat{V}_s(d\bk)\rt]
\quad \in  L^2(P,\Omega, \cF_z)\\
g=g_t(\bx,\by)&=& \int
\cF^{-1}_2\theta(\bx, \by-k^{-1}(t-z)\bk)
e^{i\bk\cdot(\bx\pm
\by/2)}
e^{ik^{-1}(t-z)|\bk|^2/2}
\hat{V}_t(d\bk)
\quad\in L^2(P,\Omega, \cF^+_t).
\eeqn
From the definition (\ref{correl})
we have
\beqn
\IE\lt[ h_s \IE_z g_t\rt]=
\IE\lt[h_s g_t\rt]
\leq \rho(t-z) \IE^{1/2}\lt[h_s^2\rt]
\IE^{1/2}\lt[g_t^2\rt].
\eeqn
Hence by  setting $s=t$ first and the
Cauchy-Schwartz inequality we have
\beqn
\IE\lt[h_s^2\rt]&\leq&\rho^2(s-z)\IE[g_t^2]\\
\IE\lt[ h_s \IE_z g_t\rt]
&\leq& \rho(t-z) \rho(s-z) \IE[g_t^2],\quad s,
\tau \geq z.
\eeqn
Therefore
\beqn
\int^\infty_z\int^\infty_z \IE[h_s g_t] ds dt
&\leq&
2\IE[g_t^2]\lt(\int^\infty_0\rho(t)dt\rt)^2
\eeqn
which, together with the integrability of $\rho(t)$,  implies a finite second
order moment of
$\tilde{V}_z$.

\commentout{***************************************************
 In terms of $\phi_p$
one has the following estimate
\beq
|\IE\lt[h_s g_t\rt]|
\leq 2^{\min{(q,2)}}\phi_p(t-z)^{1/u}
\IE^{1/(vp)}[g_t^{vp}]\IE^{1/q}[h^q_1]
\eeq
for $u, v, p, q\in [1,\infty], u^{-1}+v^{-1}=1, p^{-1}+q^{-1}=1$
and real-valued $h_s\in L^q(\Omega, \cF_z, P), g_t
\in L^{vp}(\Omega, \cF_{t}^+, P)$ (see 
\cite{EK},  Proposition~2.2).  In particular, for
$q>2, v=q/p$,
\beq
|\IE\lt[h_s g_t\rt]|
\leq 4\phi_p(t-z)^{(q-p)/q}
\IE^{1/q}[g_t^{q}]\IE^{1/q}[h^q_1],\quad
p^{-1}+q^{-1}=1
\eeq
by which, along with the  H\"older inequality, we can bound the
second moment of
$\tilde{V}_z$ as follows:

By  setting $s=t$ first and the Cauchy-Schwartz
inequality we have
\beqn
\IE\lt[h_s^2\rt]&\leq& 4\phi_p(s-z)^{(q-p)/q}
\IE^{2/q}[g_s^{q}]\\
|\IE\lt[h_s\IE_zg_t\rt]|
&\leq& 4\phi_p(s-z)^{(q-p)/(2q)}
\phi_p(t-z)^{(q-p)/(2q)}
\IE^{2/q}[g_t^{q}], \quad s,t \geq z.
\eeqn
Hence
\beqn
\int^\infty_z\int^\infty_z \IE[h_s g_t] ds dt
&=&
\int^\infty_z\int^\infty_z \IE[h_s \IE_z g_t]
ds dt\\
&\leq &8 \IE^{1/3}[g_t^6]
\lt(\int^\infty_0\phi_{6/5}^{2/5}(t) dt\rt)^2
\eeqn
 which is
finite if
$\phi_{6/5}^{2/5}(t)
$ is integrable. Note that
$\IE[g_t^6] $ in independent of
$t$ by stationarity.

When $g_t$ is almost surely bounded, the
preceding calculation with $p=1, q=\infty$ becomes
\beqn
\int^\infty_z\int^\infty_z \IE[h_s g_t] ds dt
&\leq &8  \lim_{q \to \infty}\IE^{1/q}[g_t^q]
\lt(\int^\infty_0\phi_{1}^{1/2}(t) dt\rt)^2
\eeqn
which is finite when $\phi_1^{1/2}(t)$ is integrable.
*********************************************************}

In order to bound higher order moments
in the non-Gaussian case,  one
can assume the integrability of the uniform  mixing
coefficient $\phi(t)$ \cite{EK}
\beqn
\phi(t)&=&\sup_{A\in \cF^+_{t+z}}\sup_{B\in
\cF_z\atop P(B)>0}|P(A|B)-P(A)|, \quad \forall
t\geq 0\\ &=&\sup_{A\in
\cF^+_{t+z}}\hbox{ess-sup}_\omega|P(A|\cF_z)-P(A)|
\eeqn
Then we have
\[
|P(A|\cF_z)-P(A)|\leq \phi(s-z),\quad\forall A\in \cF_{s}, \quad
s\geq z
\]
and for $p\in [1,\infty), p^{-1}+q^{-1}=1$
\beq
\IE\lt[ \lt|h_s \rt|^p\rt]\leq
2^p\phi(s-z)
\lt|\IE\lt[g_s^q\rt]\rt|^{p/q}
\eeq
(see
\cite{EK},  Proposition~2.2).
Hence the integrability of $\phi(t)$ implies  that
$\tilde{\cL}_z$ has a finite moment
of any order $p<\infty$ if $g_t$ has a finite moment
of order $q>1$.
**********************************}
\section{ Gaussian random fields}
\label{example}

By the Karhunen theorem \cite{Ka}
and  the
existence of an integrable spectral density, the random field admits
$V_z$ a moving average representation
\beq
\label{move}
V_z(\bx)=\int \Psi(z-s, \bk) W(d s, d\bk)
\eeq
where $\Psi\in L^2(\IR^{d+1})$, $W(\cdot, \cdot)$ is a
complex orthogonal random measure on $\IR^{d+1}$ such that
\[
\IE|W(\triangle)|^2=|\triangle|
\]
for all Borel sets $\triangle \subset \IR^{d+1}$. 
With
\beqn
\hat{\Psi}(\xi,\bk)&=&\frac{1}{2\pi}\int e^{-i\xi s} \Psi (s,\bk)\nn
\eeqn
 we have the following
relation between the spectral measures $\hat{V}(d\xi,d\bk)$
and $\hat{W}(d\xi,d\bk)$, on one hand, 
\beqn
\label{b}
\hat{V}(d\xi,d\bk)&=& \hat{\Psi}(\xi,\bk)\hat{W}(d\xi,d\bk)
\eeqn
 and, on other hand,
 between the spectral density $\Phi_{(\eta,\rho)}$
 and  the Fourier-transform $\hat{\Psi}$
\beqn
\Phi(\xi,\bk)&=&|\hat{\Psi}(\xi,\bk)|^2.
\eeqn

\commentout{
As a corollary of Lemma~1 and the above discussion we have 
\begin{corollary}
If $V_z$ is a Gaussian random field and its linear correlation
coefficient $r(t)$ is integrable,
then $\tilde{V}_z$ is also Gaussian and hence possesses finite
moments of all orders.
\end{corollary}
But as we have seen in Proposition~1, we need not be
concerned with the integrability of the correlation coefficient
which is a sufficient but  not necessary condition for
the square-integrability of $\tilde{V}_z$.
}

Since independence and uncorrelation are equivalent notions
for Gaussian processes, without
loss of generality, we  may take the optimal predictor
$\IE_z[V_s], s\geq z, $ to be a linear predictor, i.e., 
the orthogonal projection onto the closed linear  subspace
spanned by $\{V_t, t\leq z\}$ and write
\beq
\label{linear-span}
\IE_z[V_s(\bx)]&{=}&\int e^{i\bk\cdot\bx}
\int^z_{-\infty} C_{z,s}(\tau, \bk)\hat{V}_\tau(d\bk)
d\tau,
\quad  s\geq z\\
&=&\int e^{i\bk\cdot\bx}  \int^z_{-\infty}e^{i\xi\tau}
C_{z,s}(\tau,\bk)d\tau \hat{V}(d\xi, d\bk)\nn
\eeq
for some deterministic function 
$C_{z,s}(\tau,\bk)$ such that
\[
\int^0_{-\infty}\int^0_{-\infty} \check{\Phi}(\tau-\tau',\bk)C_{z,s}(\tau,\bk)
C_{z,s}(\tau',-\bk) d\tau d\tau'd\bk <\infty.
\]
 Indeed, the function $C_{z,s}$ satisfies the
integral equation
\beq
\label{rp}
\check{\Phi}(t-s, \bk)=\int_{-\infty}^z \check{\Phi}(t-\tau,\bk)
C_{z,s}(\tau,\bk)d\tau,\quad
\forall s\geq z\geq t, \,\,\bk\in \IR^d
\eeq
which can be obtained by averaging both sides
of (\ref{linear-span}) against $V_t(\by), t\leq z$.
Note  the  following  symmetry:
\[
\check{\Phi}(s,\bk)=\check{\Phi}(-s,\bk)=\check{\Phi}(s,-\bk),\quad 
C_{z,s}(\tau,\bk)=C_{z,s}(\tau,-\bk)
\]
analogous to (\ref{sym}).

 Hence
\beqn
\IE\lt[\IE_z[{V}_s(\bx)]\IE_z[{V}_{t}(\by)]\rt]
&=&\int
\int^z_{-\infty}\int^z_{-\infty}
\check{\Phi}(\tau-\tau',\bk)
 C_{z,s}(\tau,\bk)  C_{z,t}
(\tau',-\bk') d\tau d\tau'  
e^{i\bk(\bx-\by)}d\bk\\ &=&\int 
\check{\Phi}(t-\tau,\bk) C_{z,s}(\tau,\bk) d\tau e^{i\bk(\bx-\by)}d\bk
\eeqn
 after application of  eq. (\ref{rp}).

 \commentout{**************************
Setting $s=t$ in the above equation we see that
the spectral density of the process
$\IE_z[V_s(\bx)]$ is
given by
\beq
\IE\lt[\IE_z[\hat{V}_s(d\bk)]
\IE_z[\hat{V}_s(d\bk')]\rt]&=&
\lt[\int\Phi(\xi,\bk)d\xi\rt]
\delta(\bk+\bk') d\bk d\bk'.
\eeq
and moreover the random spectral measure 
process  $\IE_z[\hat{V}_s (d\bk)]$ has
the correlation structure
\beq
\IE\lt[\IE_z[\hat{V}_s(d\bk)]
\IE_z[\hat{V}_t(d\bk')]\rt]&=&
\lt[\int e^{i(s-t)\xi}\Phi(\xi,\bk)d\xi\rt]
\delta(\bk+\bk') d\bk d\bk'\\
&=&\check{\Phi}(s-t,\bk)\delta(\bk+\bk')d\bk d\bk',\quad s, t\geq z.
\eeq
********************************************}
Hence we have
\beqn
\IE\lt[\lt|\tilde{\cL}^\ep_z\theta\rt|^2
\rt]
&=&\int [\theta(\bx,\bp+\bq/2)-
\theta(\bx,\bp-\bq/2)][\theta(\bx,\bp+\bq/2)-
\theta(\bx,\bp-\bq/2)]\nn\\
&&\quad\times\ep^{-4}\int^\infty_0\int_0^\infty
e^{i k^{-1}(s-t)\bp\cdot\bq/\ep^{2\alpha}}
\int \check{\Phi}(\ep^{-2}(s-t),\bq) d\bq ds dt\\
&=& \int [\theta(\bx,\bp+\bq/2)-
\theta(\bx,\bp-\bq/2)]^2\nn\\
&&\quad \times\int \int^\infty_0\int_0^\infty
e^{i k^{-1}(s-t)\bp\cdot\bq\ep^{2-2\alpha}}
e^{i(s-t)\xi} \Phi(\xi, \bq) ds dt d\xi d\bq\\
&=& \int [\theta(\bx,\bp+\bq/2)-
\theta(\bx,\bp-\bq/2)]^2\nn\\
&&\quad \times
\lim_{L\to\infty}\int{|k^{-1}\bp\cdot\bq\ep^{2-2\alpha}+\xi|}^{-2}
\lt|1-e^{iL(k^{-1}\bp\cdot\bq\ep^{2-2\alpha}+\xi)}\rt|^2
\Phi(\xi, \bq) d\xi d\bq\\
&=& \int [\theta(\bx,\bp+\bq/2)-
\theta(\bx,\bp-\bq/2)]^2\nn\\
&&\times
\lim_{L\to\infty}\int{|k^{-1}\bp\cdot\bq\ep^{2-2\alpha}+\xi|}^{-2}
2\lt[1-\cos{(L(k^{-1}\bp\cdot\bq\ep^{2-2\alpha}+\xi))}\rt]
\Phi(\xi, \bq) d\xi d\bq.
\eeqn

\commentout{***********
\beqn
\lefteqn{\lan \tilde{\cL}^\ep_z\theta,
\tilde{\cL}^\ep_z\theta\ran}\\
&=&
\lan\cF^{-1}_2 \tilde{\cL^\ep}_z\theta,\cF^{-1}_2
\tilde{\cL}^\ep_z\theta\ran\\
&=&\int \ep^{-4}\int^\infty_z
\int^\infty_z\IE\lt\{\int
\cF^{-1}_2\theta(\bx,
\by-k^{-1}\ep^{-2\alpha}(s-z)\bk)
e^{i\bk\cdot(\ep^{-2\alpha}\bx\pm
\by/2)}
e^{ik^{-1}\ep^{-2\alpha}(s-z)|\bk|^2/2}
\IE^\ep_z\lt[
\hat{V}^\ep_s(d\bk)\rt]
\right.\\
&&\lt. \int
\cF^{-1}_2\theta(\bx,
\by-k^{-1}\ep^{-2\alpha}(t-z)\bk')
e^{-i\bk'\cdot(\epal\bx\pm
\by/2)}
e^{-ik^{-1}\ep^{-2\alpha}(t-z)|\bk'|^2/2}
\overline{\hat{V}^\ep_t(d\bk')}
\rt\}\,\, ds dt  d\bx
d\by\\
&=&\int \ep^{-4}\int^\infty_0
\int^\infty_0\int
\cF^{-1}_2\theta(\bx,
\by-k^{-1}\ep^{-2\alpha}s\bk)
\cF^{-1}_2\theta(\bx,
\by-k^{-1}\ep^{-2\alpha}t\bk)\\
&&\quad\quad
e^{ik^{-1}\ep^{-2\alpha}(s-t)|\bk|^2/2}
\int e^{i\ep^{-2\alpha}(s-t)\xi}\Phi(\xi,\bk) d\xi
d\bk ds dt  d\bx d\by\\
&=&\int \int^\infty_{0}
\int^\infty_{0}\int
\cF^{-1}_2\theta(\bx,
\by-k^{-1}\ep^{2-2\alpha}s\bk)
\cF^{-1}_2\theta(\bx,
\by-k^{-1}\ep^{2-2\alpha}t\bk)\\
&&\quad\quad
e^{ik^{-1}\ep^{2-2\alpha}(s-t)|\bk|^2/2}
\int e^{i\ep^{-2\alpha}(s-t)\xi}\Phi(\xi,\bk) d\xi
d\bk ds dt  d\bx d\by
\eeqn
}

\end{appendix}
*******************}

\end{document}